\documentclass[aps,floatfix,showpacs,prb]{revtex4}
\usepackage{graphicx,amsfonts,amssymb,enumerate,color}
\usepackage{natbib}

\newcommand{\be}{\begin{equation}}
\newcommand{\ee}{\end{equation}}
\newcommand{\bea}{\begin{eqnarray}}
\newcommand{\eea}{\end{eqnarray}}
\newcommand{\beb}{\begin{eqnarray*}}
\newcommand{\eeb}{\end{eqnarray*}}

\newcommand{\LLletter}{\ensuremath{n}}
\newcommand{\aspectratio}{\ensuremath{\mathrm{AR}}}

\begin{document}
\title{The Quantum Hall ferroelectric helix in bilayer graphene}

\author{Thierry Jolicoeur$^1$}
\author{Csaba T\H{o}ke$^{2,3}$}
\author{Inti Sodemann$^4$}

\affiliation{$^1$Laboratoire de Physique Th\'eorique et Mod\`eles statistiques, 
CNRS, Universit\'e Paris-Sud, Universit\'e Paris-Saclay, 91405 Orsay, France}

\affiliation{$^2$Department of Theoretical Physics, Budapest University of Technology and Economics, 
Budafoki \'ut 8, H-1111 Budapest, Hungary}

\affiliation{$^3$BME ``Momentum'' Exotic Quantum Phases Research Group, 1111 Budapest, Budafoki \'ut 8, Hungary}

\affiliation{$^4$Max-Planck Institute for the Physics of Complex Systems, D-01187 Dresden, Germany}

\date{December, 2018}
\begin{abstract}

We re-examine the nature of the ground states of bilayer graphene at odd integer filling factors within a simplified model of nearly degenerate $n=0$ and $n=1$ Landau levels. Previous Hartree-Fock studies have found that ferroelectric states with orbital coherence can be stabilized by tuning the orbital splitting between these levels. These studies indicated that, in addition to a uniform ferroelectric state, a helical ferroelectric phase with spontaneously broken translational symmetry is possible. By performing exact diagonalization on the torus, we argue that the system does not have a uniform coherent state but instead transitions directly from the uniform incoherent state into the ferroelectric helical phase. We argue that there is a realistic prospect to stabilize the helical ferroelectric state in bilayer graphene by tuning the interlayer electric field in a model that includes all single particle corrections to its zero energy eight-fold multiplet of Landau levels.
\end{abstract}
\maketitle

\section{Introduction}

\begin{figure}[t]
\centering
 \includegraphics[width=0.5\columnwidth]{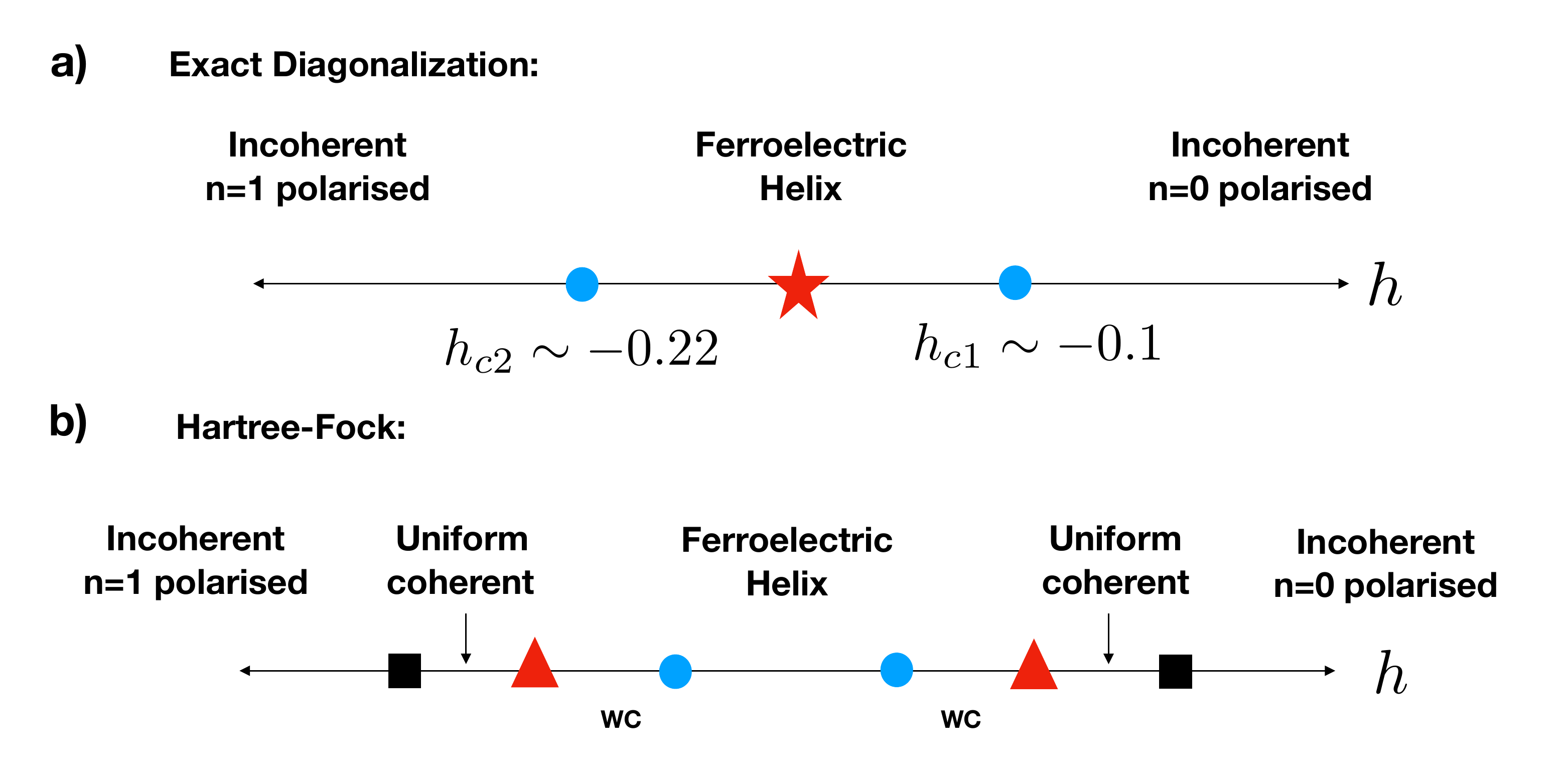}
 \caption{Summary of the phases observed in the orbitally mixed $n=0/1$ as a function of the orbital single particle splitting $h$. (a) 
 The exact diagonalization results of the current study indicate the presence of only two phases: an incoherent adiabatically connected to the 
 maximally polarized states, and a ferroelectric helical state. The red star indicates the orbital bias at which particle hole symmetry is 
 achieved for which the system has equal population of the $n=0/1$ orbitals. (b) The Hartree-Fock studies find two other phases: the uniform 
 orbitally coherent ferroelectric and a Wigner crystal (WC), see e.g. Ref~.(\onlinecite{Cote11}) for a detailed estimate of the boundaries.}
 \label{cartoon}
\end{figure}

The last two decades have witnessed a rapid and formidable increase in the richness of the quantum Hall physics of monolayer and bilayer 
graphene~\cite{Novoselov05,Novoselov06,Du09,Bolotin09,Dean11,Feldman12,Smet12,Young12,CastroNeto09,Abanin09,Abanin13,Toke07,Ki13,Papic11-1,Papic11-2}. 
In particular, bilayer graphene (BLG) possesses a unique zeroth Landau level manifold which features two nearly degenerate cyclotron orbital 
degrees of freedom with $n=0$ and $n=1$ character~\cite{McCann06}, in addition to the two spins and two valleys making up a manifold with a 
total of eight levels. The interplay of strong Coulomb interactions, that stabilize integer quantum Hall states via quantum Hall ferromagnetism~\cite{YSM}, 
and the single particle splitting terms, can lead to an intricate variety of coherent states in this system~\cite{Barlas08,Barlas10,Cote10,Cote11,Lambert,Knothe}.
A notoriously interesting possibility, pointed out in Ref.~\onlinecite{Barlas08}, is the coherence between the states with different orbital 
character $\LLletter=0/\LLletter=1$. This coherence breaks spontaneously the real space inversion symmetry, resulting in the formation of a 
type of quantum Hall ferroelectric state, which is expected to have a linearly dispersing Goldstone mode (this form of quantum Hall 
ferroelectric state is distinct from that proposed in Ref.~\onlinecite{Zheng} which results from spontaneous valley polarization). 
Following these initial studies, it was later argued based on Hartree-Fock theory~\cite{Cote11}, that an analogue of Dzyaloshinskii-Moriya 
interaction allowed by the breaking of inversion symmetry, can drive the softening of this Goldstone mode at a finite wave-vector, leading to 
the formation of ferroelectric helical state. The phase diagram as a function of the single-particle splitting of the $\LLletter=0/\LLletter=1$ 
orbitals obtained from Hartree-Fock is depicted in Fig.\ref{cartoon}(b), which additionally features a Wigner crystal state.

To this date there are very few studies incorporating correlation effects beyond Hartree-Fock on this problem. One exact diagonalization 
study~\cite{AbaninPapic} restricted itself to the case of zero single particle splitting between $n=0$ and $n=1$ orbitals, and found that 
in such case the ground state is a trivial integer quantum Hall state adiabatically connected to the fully polarized state into the $n=0$ 
orbital with a full gap to all excitations. As we will see, however the interesting ferroelectric states predicted by Hartree-Fock theory 
tend to occur at negative splitting when the $n=1$ orbital is energetically favored, explaining why they were not observed in Ref.~(\onlinecite{AbaninPapic}).

In this paper we concentrate on phases with full spin and valley polarization and just keep the two orbital degrees of freedom $\LLletter=0$ and $\LLletter=1$, 
and restrict to total filling factor $\nu=1$. Using the torus geometry we diagonalize exactly the Hamiltonian for up to 14 electrons, which allows us 
to obtain information on ground states and some of the low-lying excited states. We pay special attention to the role of a nontrivial particle-hole 
symmetry, identified in Ref.~(\onlinecite{Shizuya12}), which maps states with filling $\nu$ onto states with $2-\nu$ in the two orbital manifold. 
Remarkably, this symmetry acts non-trivially on the bare single particle-splitting between the $\LLletter=0$ and $\LLletter=1$ orbitals, and for a 
Hamiltonian to be particle-hole invariant under this symmetry it must have a negative splitting favoring the $\LLletter=1$ orbital.

We will show that the simple polarized incoherent phase observed in Ref.(\onlinecite{AbaninPapic}) can be captured by perturbation theory and show 
that its excited states can be reproduced by time-dependent Hartree-Fock theory. We also find that for some range of orbital splitting there is 
evidence for a phase with broken translation symmetry that we identify as the helical ferroelectric phase seen in HF calculations. However, we find 
no evidence for a spatially uniform orbitally coherent state, because we do not observe any translationally invariant phase with its characteristic 
Goldstone mode. We depict the approximate phase diagram resulting from our study in Fig.\ref{cartoon}(a).
We will also show that there is a realistic prospect to realise the ferroelectric helical state in BLG. Recent experiments have achieved a detailed 
understanding and a remarkable degree of control over the single particle splittings in the zeroth Landau level manifold of BLG~\cite{Hunt,Andrea2017,Dean}. 
Because the valley degree of freedom is locked to the layer index in the zeroth Landau level of BLG, this degree of freedom can be easily controlled by 
applying an interlayer bias. The spin splitting on the other hand can be controlled with in-plane fields. It is therefore, possible to achieve the 
conditions in which the system is valley and spin polarized and the relevant active degrees of freedom are the $\LLletter=0$ and $\LLletter=1$ orbitals. 
The splitting between the $n=0$ and $n=1$ levels is intimately related to hopping terms that break particle-hole invariance in bilayer graphene~\cite{Jeil}. 
We will also show, however, that there is a way to experimentally control the splitting between the $\LLletter=0$ and $\LLletter=1$ orbitals by tuning 
the interlayer bias to sufficiently large values.

Our paper is organized as follows, in section II we discuss the realistic band structure of BLG and show that the level crossing between 
$\LLletter=0/\LLletter=1$ lies within a realistic range of parameters. In section III we explain the peculiar particle-hole symmetry of 
the model obtained by assuming full spin and valley polarization. Section IV contains some definitions of finite-size torus wavefunctions. 
Section V summarize the Hartree-Fock treatment. In section VI we give results of exact diagonalization studies. Section VII is devoted to 
the incoherent phase and its excitations. Section VIII contains our findings about the broken translation symmetry phase. Our conclusions 
are presented in section IX.

\section{Band structure of bilayer graphene}

The four band model containing all leading corrections to bilayer graphene's Hamiltonian for a single spin and valley in the presence of a magnetic field has the form~\cite{Jeil}~:
\be
H_{4b}=\omega_0 \left(
\begin{array}{cccc}
 \frac{u}{2  \omega_0} & a^\dagger & - \frac{\gamma_4}{\gamma_0} a^\dagger  & - \frac{\gamma_3}{\gamma_0} a  \\
 a & \frac{\Delta'+ u}{2  \omega_0} &  \frac{\gamma_1}{\omega_0} & - \frac{\gamma_4}{\gamma_0} a^\dagger \\
- \frac{\gamma_4}{\gamma_0}  a &  \frac{\gamma_1}{\omega_0} &  \frac{\Delta'-u}{2  \omega_0} & a^\dagger \\
 - \frac{\gamma_3}{\gamma_0} a^\dagger  & - \frac{\gamma_4}{\gamma_0} a  & a & - \frac{u}{2  \omega_0} \\
\end{array}
\right).
\ee
\noindent Here $\omega_0=\sqrt{2} v_0/\ell \approx 30.6{\rm meV} \sqrt{{\rm B[T]}}$, 
with the magnetic length $\ell =\sqrt{\hbar c /eB}$,
$ \gamma_0\approx 2.61$eV, $ \gamma_1\approx 361$meV, $\gamma_3\approx 283$meV, $\gamma_4\approx 138$meV, $\Delta'\approx 15$meV, and $u$ is the energy difference between top and bottom layers controlled by a perpendicular electric field. This Hamiltonian has been successfully employed in detailed modeling of Landau levels for integer~\cite{Hunt} and fractional quantum Hall states recently~\cite{Andrea2017}.

\begin{figure}[t]
\centering
 \includegraphics[width=0.5\columnwidth]{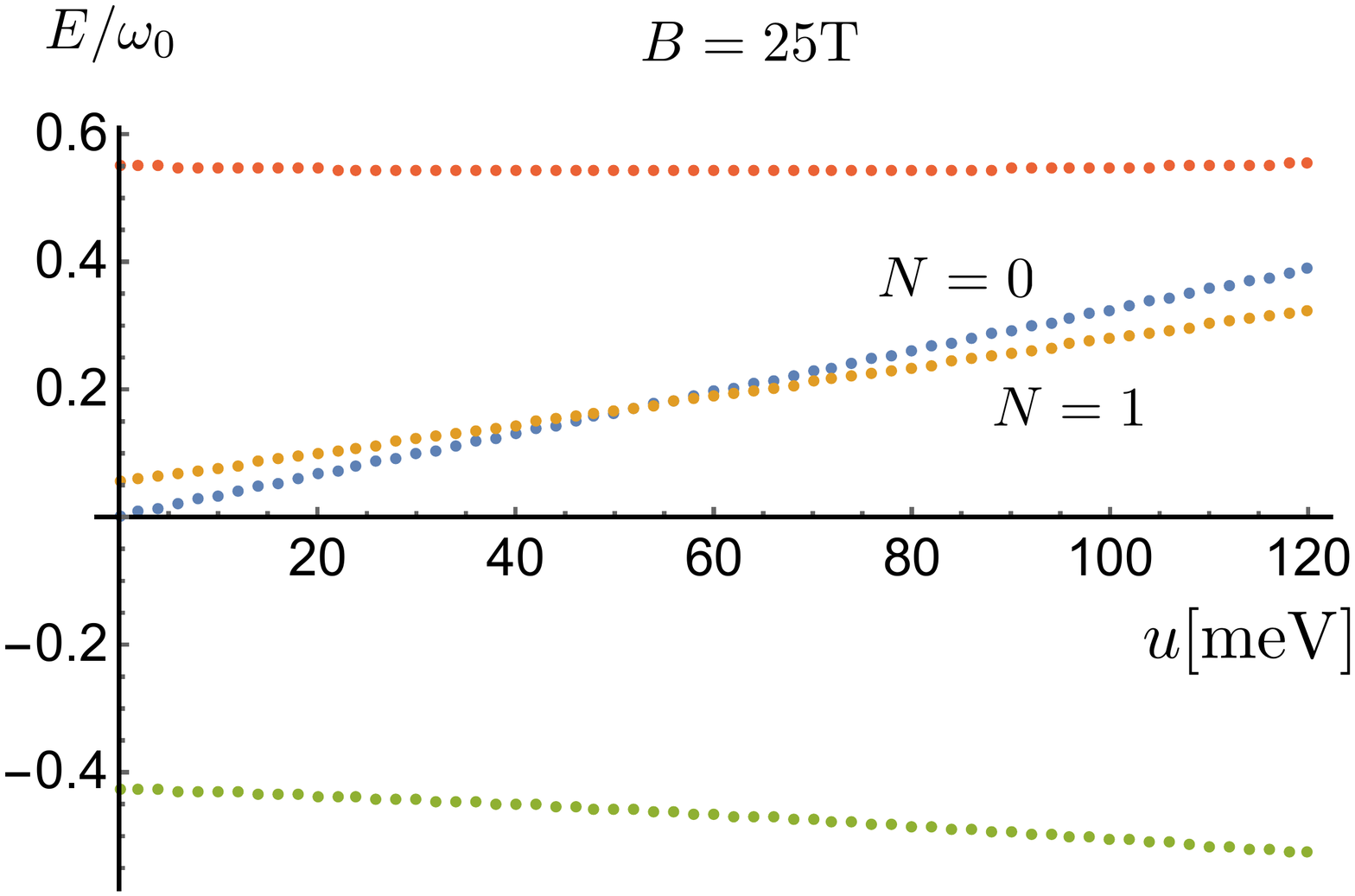}
 \caption{Landau levels for a single spin-valley in BLG as a function of interlayer bias. The bias can tune a level crossing between $\LLletter=0$ and $\LLletter=1$. There is a sizable energy separation from higher Landau levels throughout justifying the projection into the two $\LLletter=0$ and $\LLletter=1$ LLs.}
 \label{LLs}
\end{figure}

\begin{figure}[t]
\centering
 \includegraphics[width=0.5\columnwidth]{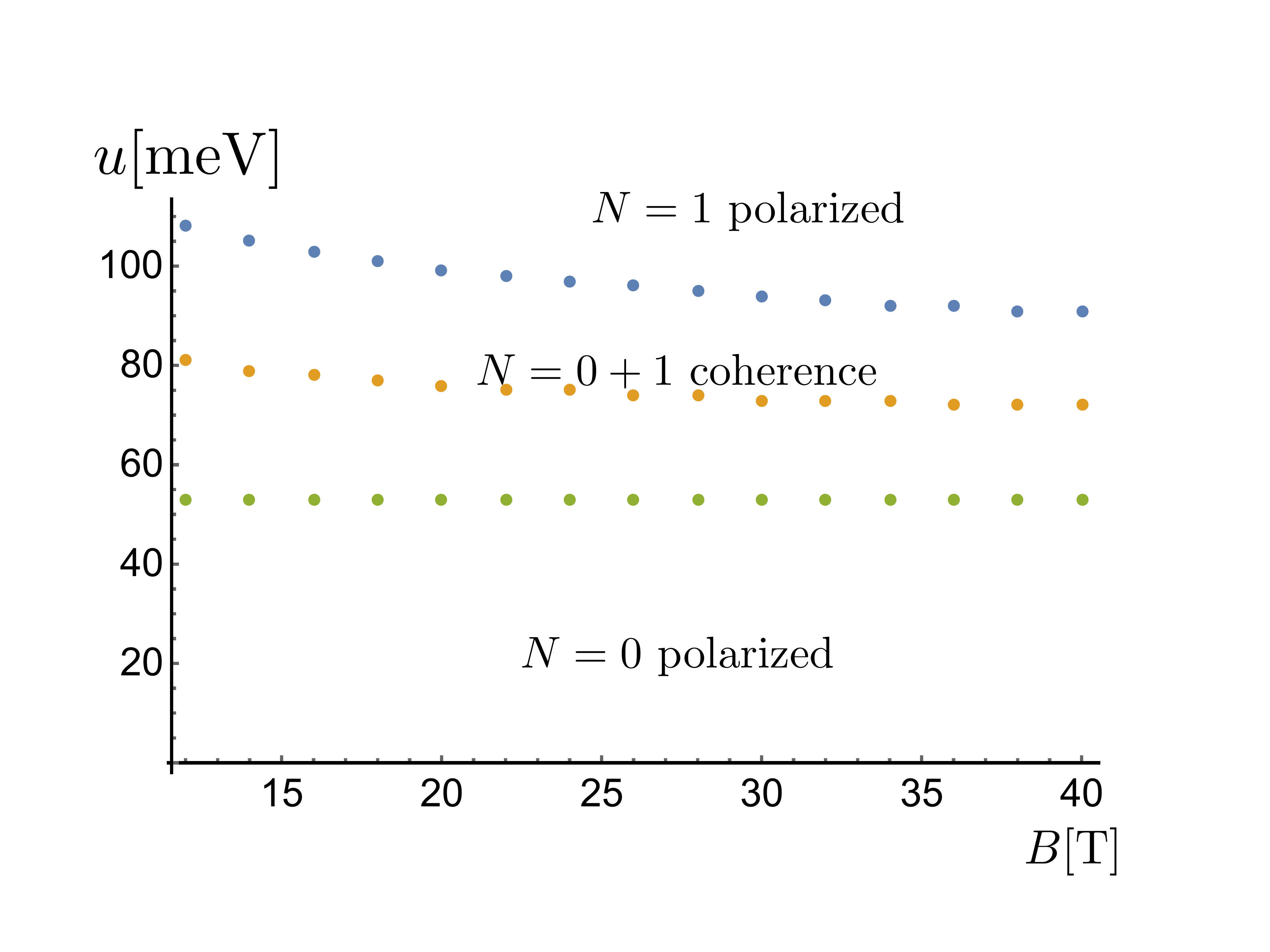}
 \caption{Expected schematic phase diagram as a function of interlayer bias $u$ and perpendicular magnetic field. Identification of phases come Hartree-Fock calculations.
 The green dots are where the bare single particle energies of $\LLletter=0$ and $\LLletter=1$ become degenerate. The yellow dots lie where approximate particle symmetry is expected in the presence of Coulomb interactions and the blue dots are the expected location at which the system polarizes into $\LLletter=1$. We assume that the bilayer graphene sample is on top of a hexagonal boron nitride substrate with screening constant $\epsilon\approx 6.6$ as in ref.(\onlinecite{Andrea2017}).}
 \label{PhaseD}
\end{figure}

\begin{figure}[t]
\centering
 \includegraphics[width=0.8\columnwidth]{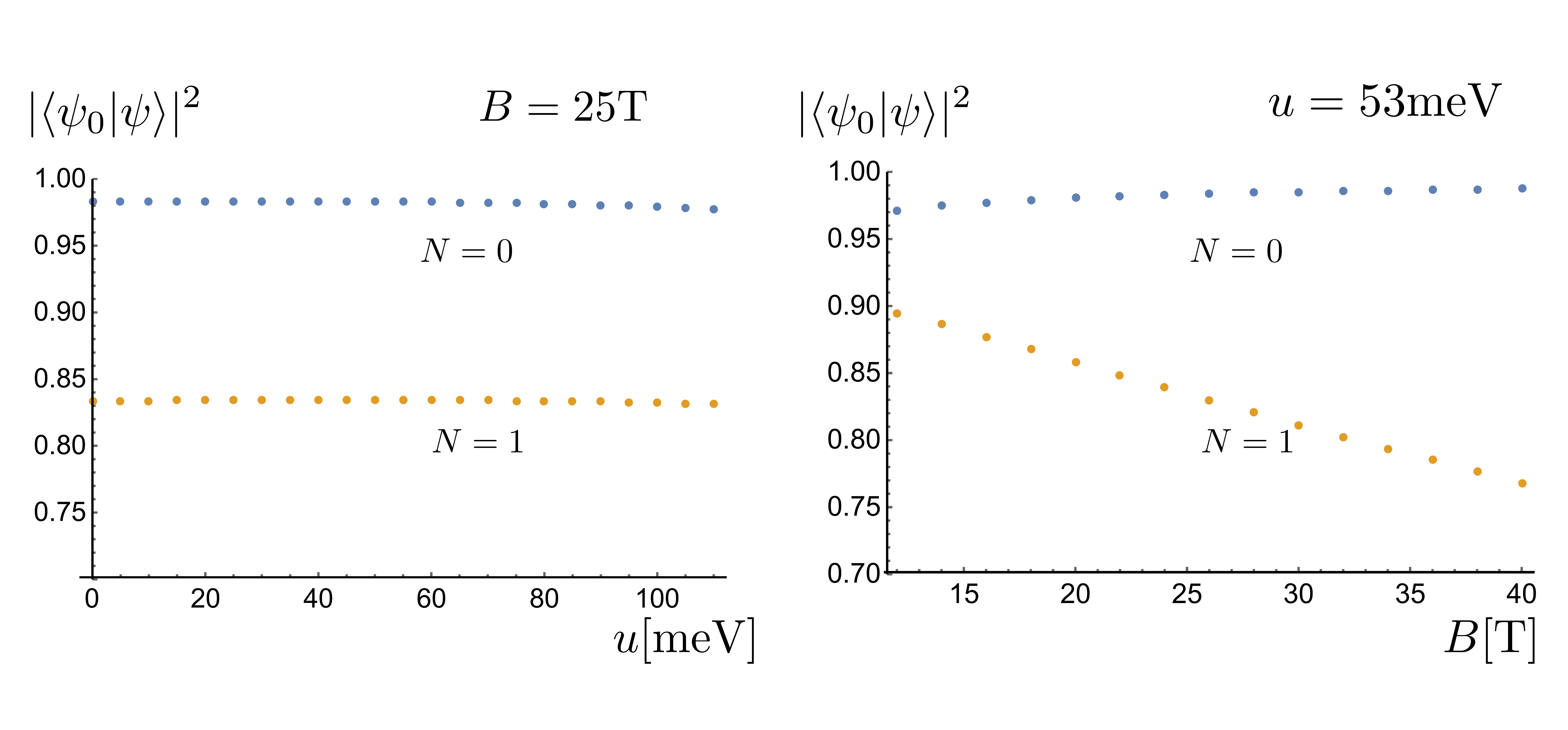}
 \caption{Overlaps between the ideal two-band model Landau level wavefunctions and the four-band model wavefunctions as a function of interlayer bias and magnetic field.}
 \label{overlap}
\end{figure}

In this Hamiltonian the terms $\Delta'$ and $\gamma_4$ break particle-hole symmetry and induce a splitting between the $\LLletter=0$ and $\LLletter=1$ LLs at zero interlayer bias $u$ which favors $\LLletter=0$. This renders the ground state trivially polarized into $\LLletter=0$ at $u=0$, and therefore under normal conditions one would not expect the physics that we will describe in this paper to appear. However, as illustrated in Fig.~\ref{LLs}, the interlayer bias can be used to control the splitting between these levels and successfully induce a level crossing making the $\LLletter=1$ lower in energy as we desire. The value required to achieve this crossing is independent of the magnetic field and, neglecting the trigonal warping term ($\gamma_3 \rightarrow 0$), can be estimated to be~:
\be
u_0=\frac{2 \gamma _4 \gamma _1}{\gamma _0}+\Delta'  \left(1+\left(\frac{\gamma _4}{\gamma _0}\right)^2\right)\approx  53.2{\rm meV}.
\ee
which should be within experimental reach. In Fig.~\ref{PhaseD} we show this value as green dots (including the trigonal warping term). At this value one expects the boundary between the state fully polarized into $\LLletter=0$ and the coherent states. The orange dots are the required values in order to overcome the splitting produced by the exchange interactions with the vacuum (analogous to the well-known Lamb shift in atomic physics)\cite{Shizuya12} for bare Coulomb interactions so that the system has approximate particle-hole symmetry.
The blue dots depict the expected Hartree-Fock boundary between coherent and fully $\LLletter=1$ polarized states \cite{Cote11}.
Taking bare Coulomb interactions, these two boundaries correspond to the single particle energy splitting given respectively by:
\begin{eqnarray*}
  E_1-E_0=\epsilon_{\mathrm{ Lamb}}&=&\frac{e^2}{\epsilon \ell}\frac{1}{16}\sqrt{\frac{\pi}{2}},\\
  E_1-E_0=\epsilon_{\mathrm{ orb. pol.}}&=&\frac{e^2}{\epsilon \ell}\frac{1}{8}\sqrt{\frac{\pi}{2}}.
\end{eqnarray*}
Finally, in Fig.~\ref{overlap} we illustrate the overlaps of the LL levels in the full four-band model with the idealized two-band model as a function of the interlayer bias and field. We see an excellent overlap for $\LLletter=0$ and a good overlap for $\LLletter=1$. The latter is known to decrease as the magnetic field increases~\cite{Andrea2017}, but as we see there is no substantial change as a function of interlayer bias.

\section{Particle-hole symmetry between $\LLletter=0$ and $\LLletter=1$ Landau levels}

Consider the two-band model of bilayer graphene~:
\be
\mathcal{H}_K =\left(
\begin{matrix}
{0 & -(\Pi_x-i\Pi_y)^2/2m^* \cr
 -(\Pi_x+i\Pi_y)^2/2m^* & 0 }
\end{matrix}\right)
\label{H_K}
\ee
where $m^*$ is an effective mass and $\Pi_{x,y}=p_{x,y}+eA_{x,y}/c$. This is the Hamiltonian concerning valley $K$ and spin is polarized. The single-particle spectrum has then
two zero-energy states~:
\be
E_{0K}=E_{1K}=0
\ee
and the remainder of the spectrum depends upon the magnetic field~:
\be
E_{nK}=\mathrm{sign}(n)\omega_c \sqrt{|n|(|n|-1)},\quad |n| \geq 2 ,
\ee
where $n$ is an integer positive or negative and $\omega_c=eB/m^*$. 
The corresponding eigenstates are given by~:
\be
\Phi_n=\left(\begin{matrix}
{\phi_n  \cr 0 }\end{matrix}\right),\quad \LLletter=0,1
\ee
\be
\Phi_n=\frac{1}{\sqrt{2}}\left(\begin{matrix}
{\phi_{|n|}  \cr  \phi_{|n|-2}}\end{matrix}\right),\quad n\leq -2,\quad
\Phi_n=\frac{1}{\sqrt{2}}\left(\begin{matrix}
{-\phi_{|n|}  \cr  \phi_{|n|-2}}\end{matrix}\right),\quad n\geq +2,
\ee
where $\phi_n$ are standard cyclotron eigenstates for particles with parabolic
dispersion relation. We now consider two-body interactions written in second
quantization~:
\be
\mathcal{V}=\frac{1}{2}\sum_{1234}V_{1234}\,\,
c^\dag_{n_1,j_1}c^\dag_{n_2,j_2}c_{n_3,j_3}c_{n_4,j_4}
\label{Voperator}
\ee
with Landau level indices $n_i=0,1$ and guiding center coordinates $j_i$
and the big indices $\{1234\}$ are a short-hand for the combined indices.
We now use the particle-hole conjugation operator $\mathcal{C}$ which is
defined by~:
\be
\mathcal{C}^\dag c^\dag_{nj}\mathcal{C}=c_{nj}.
\ee
When acting on Eq.(\ref{Voperator}) we find~:
\be
\mathcal{C}^\dag \mathcal{V}\mathcal{C}=
\mathcal{V}+\sum_{12M}c^\dag_1 c_2 (V_{1M2M}+V_{M2M1}-V_{2MM1}-V_{M21M})
+{\rm const}
\ee
The particle-hole transformation generates an additional one-body term
which can written as ~:
\be
\Delta_0\sum_m c^\dag_{0m}c_{0m}+
\Delta_1\sum_m c^\dag_{1m}c_{1m},
\label{split}
\ee
The constants $\Delta_{0,1}$ are geometry-dependent and they are \textit{distinct}
$\Delta_0\neq\Delta_1$. So even if we start with exactly degenerate $\LLletter=0$ and $\LLletter=1$ LLs
the particle-hole symmetry leads to a nontrivial splitting between them. If we now focus
on the state with total filling factor $\nu=1$ then the particle-symmetry above implies
that the full spectrum of the Hamiltonian Eq.(\ref{Voperator}) should be  the same
after addition of the one-body splitting Eq.(\ref{split}).

\section{Torus definitions}
\label{sec:torus}

We consider a torus that is obtained by periodic boundary conditions applied to a rectangular domain
spanned by vectors $\mathbf L_1=L_x\mathbf{\hat x}$ and $\mathbf L_2=L_y\mathbf{\hat y}$.
The aspect ratio is defined as~:
\begin{equation}
  \aspectratio=\frac{L_x}{L_y}.
\end{equation}
Periodic boundary conditions are imposed by magnetic translations \cite{Zak64}
$t(\mathbf L)=\exp\left(\frac{i}{\hbar}\mathbf L\cdot\Pi-i\frac{\mathbf{\hat z}\cdot(\mathbf L\times\mathbf r)}{\ell^2}\right)$.
Using the Landau gauge $\mathbf A=Bx\mathbf{\hat y}$ for in this study,
these act as $t(\mathbf L)\psi(\mathbf r) = \exp(\frac{iy\mathbf{\hat x}\cdot\mathbf L}{\ell^2})\psi(\mathbf r+\mathbf L)$.
The boundary conditions read $t(\mathbf L_{1,2})\psi(\mathbf r)=\psi(\mathbf r)$.
The two conditions are compatible only if the rectangle is pierced by an integral number of flux quanta~:
\begin{equation}
N_\phi=\frac{\left|\mathbf L_1\times\mathbf L_2\right|}{2\pi\ell^2}=\frac{L_x L_y}{2\pi\ell^2}.
\end{equation}
The $\LLletter=0$ states can be written as~\cite{Haidekker18}~:
\begin{equation}
  \label{eq:eta0}
  \eta_{0m}(z)=\frac{1}{\sqrt{\ell L_2\sqrt\pi}}\vartheta\left[\begin{array}{c} m/N_\phi \\ 0\end{array}\right]
  \left(\frac{i\pi N_\phi z}{L_2}\Big|N_\phi i\right)e^{-\frac{x^2}{2\ell^2}},
\end{equation}
where $m=0,1,\dots,(N_\phi-1)$, $z=x+iy$, and we have used Jacobi elliptic functions with characteristics~\cite{Mumford87}~:
\begin{equation}
  \label{eq:theta}
  \vartheta\left[\begin{array}{c} a \\ b \end{array}\right](z|\tau) = \sum_{n=-\infty}^\infty e^{i\pi\tau(n+a)^2+2i(n+a)(z+b\pi)}.
\end{equation}
Higher Landau orbitals are obtained using the LL raising operator, which in our gauge takes the form~:
$\hat a^\dag = \frac{\ell}{\sqrt2}\left(i\partial_x+\partial_y-\frac{ix}{\ell^2}\right)= \frac{\ell}{\sqrt2}\left(2i\partial_z-\frac{ix}{\ell^2}\right)$.
Thus we have~:
\begin{eqnarray}
  \label{eq:eta1}
  \eta_{n,j}(z)&=&\frac{(\hat a^\dag)^n}{\sqrt{n!}}\eta_{n0}(z)=\\
  &=&\frac{1}{(L_y\sqrt{\pi})^{1/2}}
  \sum_k H_n\left(\frac{x}{\ell}-\frac{kL_x}{\ell}-\frac{2\pi j\ell}{L_y}\right)
  \exp\left(-i\left(\frac{kL_x}{\ell^2}+\frac{2\pi j}{L_y}\right)y
  -\frac{1}{2\ell^2}\left(x-kL_x-\frac{2\pi j\ell^2}{L_y}\right)^2\right),\nonumber
\end{eqnarray}
where $H_n$ is a a Hermite polynomial.
We note that $\eta_{nm}(z)$ is normalized for the principal domain of the torus~:
\begin{equation}
  \int_0^{L_y}dy\int_{0}^{L_x}dx \eta^\ast_{n'm'}(x+iy)\eta_{nm}(x+iy)=\delta_{nn'}\delta_{mm'}.
\end{equation}


\section{Hartree-Fock treatment}

The difference of the spatial profile of the $\LLletter=0$ and $\LLletter=1$ Landau states allows for
nontrivial electric dipole structures when both of these states are relevant to the ordering
in a partially filled Landau band.
C\^ot\'e, Fouquet and Luo \cite{Cote11} have elaborated the Hartree-Fock mean-field theory for such systems.
Treating the energy splitting $h$ between the $\LLletter=1$ and the $\LLletter=0$ orbitals as a parameter,
the mean-field energy per particle reads, in units of $e^2/(\epsilon\ell)$ and having the thermodynamic limit in mind, as~:
\begin{eqnarray}
  \label{eq:cotehf}
  \frac{E_\textrm{\scriptsize HF}}{N}&=&-\frac{11}{32}\sqrt{\frac{\pi}{2}}-h\left\langle\rho_z(\mathbf q=0)\right\rangle\\
  &&+\frac{1}{2}\sum_{\mathbf q}a(q)\left(\left\langle\rho_x(-\mathbf q)\right\rangle\left\langle\rho_x(\mathbf q)\right\rangle + \left\langle\rho_y(-\mathbf q)\right\rangle\left\langle\rho_y(\mathbf q)\right\rangle\right)+\nonumber\\
  &&+\frac{1}{2}\sum_{\mathbf q}b(q)
  \left(\begin{array}{cc} \left\langle\rho_x(-\mathbf q)\right\rangle & \left\langle\rho_y(-\mathbf q)\right\rangle \end{array}\right)
  \left(\begin{array}{cc} \cos(2\varphi_q) & \sin(2\varphi_q) \\ \sin(2\varphi_q) & -\cos(2\varphi_q)\end{array}\right)
  \left(\begin{array}{cc} \left\langle\rho_x(\mathbf q)\right\rangle \\ \left\langle\rho_y(\mathbf q)\right\rangle \end{array}\right)+\nonumber\\
  &&+\frac{1}{2}\sum_{\mathbf q}c(q)\left\langle\rho_z(-\mathbf q)\right\rangle\left\langle\rho_z(\mathbf q)\right\rangle+\nonumber\\
  &&+\frac{i}{4}\sum_{\mathbf q}d(q)(\mathbf{\hat z}\times\mathbf{\hat q})\cdot(\left\langle\vec\rho(-\mathbf q)\right\rangle\times\left\langle\vec\rho(\mathbf q)\right\rangle),\nonumber
\end{eqnarray}
where the pseudospin density operator $\rho(\mathbf q)$ is written as 
$\rho_x(\mathbf q)=-(\rho_{1,0}(\mathbf q) + \rho_{0,1}(\mathbf q))/2$,
$\rho_y(\mathbf q)=(\rho_{1,0}(\mathbf q) - \rho_{0,1}(\mathbf q))/2i$,
$\rho_z(\mathbf q)=(\rho_{0,0}(\mathbf q) - \rho_{1,1}(\mathbf q))/2$, with
\begin{equation}
\rho_{n,n'}(\mathbf q)=\frac{1}{N_\phi}\sum_X e^{-iq_xx +iq_xq_y\ell^2/2}c^\dag_{nX}c_{n',X-q_y\ell^2};
\end{equation}
$\varphi_q$ is the angle of vector $\mathbf q$, and we have used real-valued functions that eventually follow from the matrix
elements of the Coulomb interaction, defined in Ref.~\onlinecite{Cote11}, which can be written in closed form as
\begin{eqnarray*}
  a(q)
  &=&q\ell e^{-q^2\ell^2/2}
  - \sqrt{\frac{\pi}{2}}\left(1+q^2\ell^2/2\right)e^{-q^2\ell^2/4} I_1\left(q^2\ell^2/4\right)
  - \sqrt{\frac{\pi}{8}}q^2\ell^2 e^{-q^2\ell^2/4} I_1\left(q^2\ell^2/4\right),\\
  b(q)
  &=&q\ell e^{-q^2\ell^2/2} - \sqrt{\frac{\pi}{8}}e^{-q^2\ell^2/4} I_0\left(q^2\ell^2/4\right)+
   \sqrt{\frac{\pi}{2}} \left(1+q^2\ell^2/2\right) e^{-q^2\ell^2/4} I_1\left(q^2\ell^2/4\right),\\
  c(q)
  &=&\frac{q^3\ell^3}{4} e^{-q^2\ell^2/2}
  - \sqrt{2\pi}e^{-q^2\ell^2/4} \left(\frac{3}{8}+\frac{q^2\ell^2}{8}+\frac{q^4\ell^4}{16}\right)I_0\left(q^2\ell^2/4\right)+
  \sqrt{2\pi}e^{-q^2\ell^2/4} \left(\frac{q^4\ell^4}{16}-\frac{q^2\ell^2}{4}\right)I_1\left(q^2\ell^2/4\right),\\
  d(q)&=&
  \sqrt2 q^2\ell^2 e^{-q^2\ell^2/2}
  + \sqrt{\frac{\pi}{32}} q^3\ell^3 e^{-q^2\ell^2/4} I_0\left(q^2\ell^2/4\right)
  - \sqrt{\frac{\pi}{32}} q(3+q^2\ell^2)\ell e^{-q^2\ell^2/4} I_1\left(q^2\ell^2/4\right).
\end{eqnarray*}
Here, $h$ can be related to the interlayer bias  $u$
but we prefer to handle it as a parameter throughout this study.
The $a(q)$ and $c(q)$ terms are anisotropic exchange coupling between pseudospins, reminiscent of the XXZ model;
the $b(q)$ term is basically the dipole-dipole electrostatic interaction;
and the $d(q)$ term is a Dzyaloshinskii-Moriya type interaction between pseudospins.
We emphasize that even though the above formulas refer to $N_\phi$ flux quanta, the $N_\phi\to\infty$
limit is implied; several assumptions in the analysis based on Eq.~(\ref{eq:cotehf}) are applicable only this case.
In the phase diagram based on Eq.~(\ref{eq:cotehf}) the region around the particle-hole symmetric
value of $h=0$ the system exhibits pseudospin helical state, while a uniform liquid is expected outside of this region.
The latter uniform state may exhibit orbital phase coherence with the ensuing ferroelectric dipole ordering.

\begin{figure}[htbp]
\begin{center}
\includegraphics[width=0.33\textwidth, keepaspectratio]{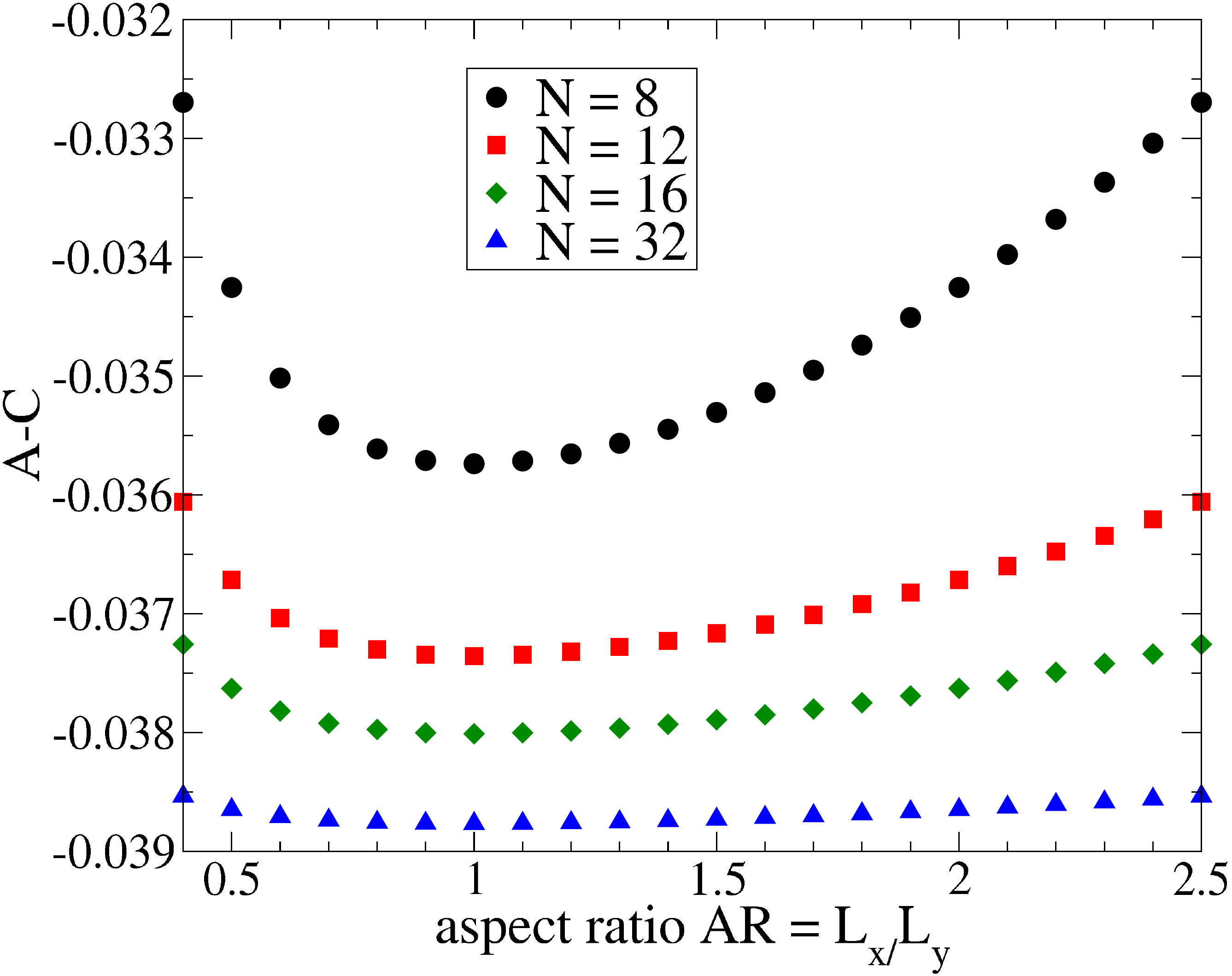}
\includegraphics[width=0.33\textwidth, keepaspectratio]{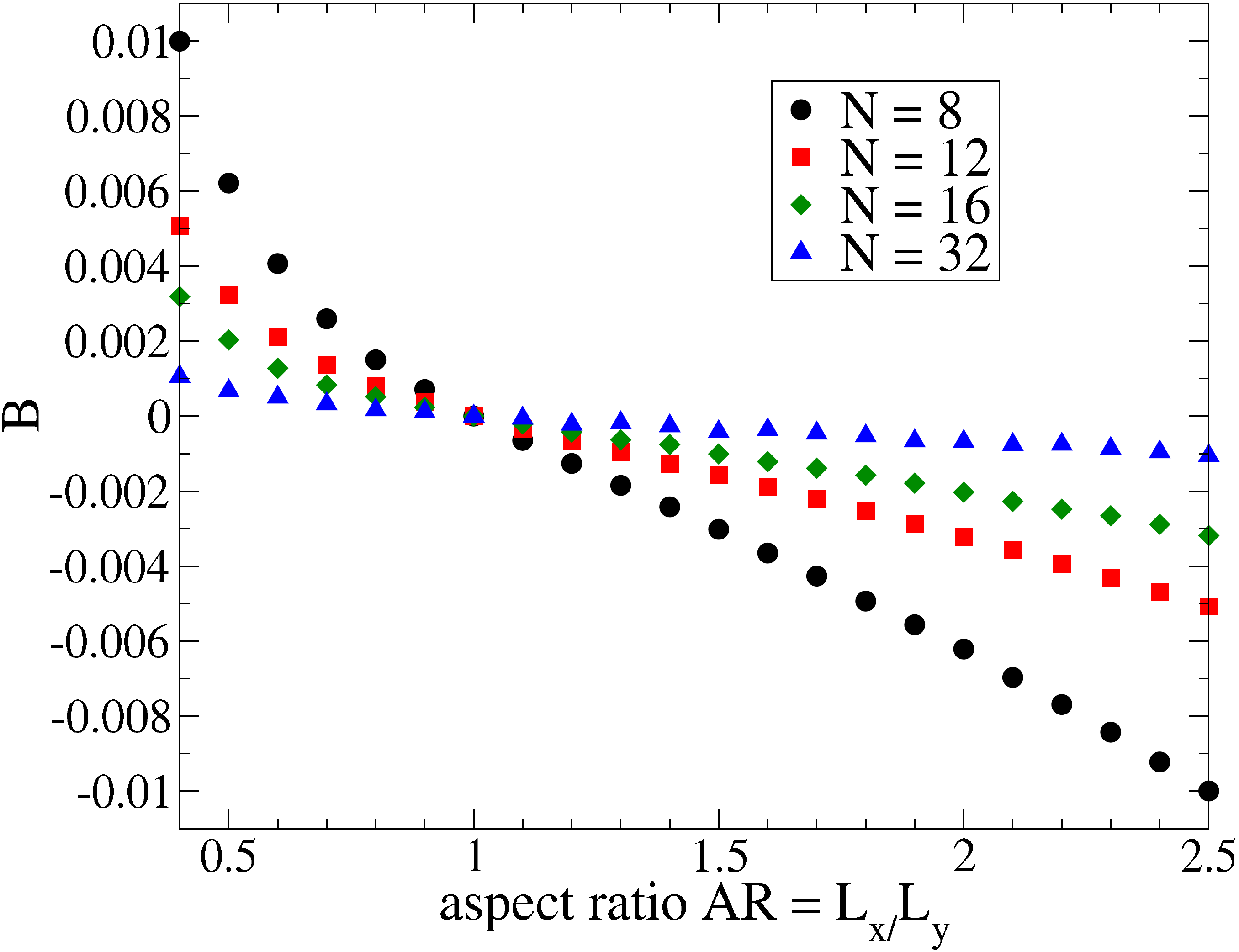}
\includegraphics[width=0.33\textwidth, keepaspectratio]{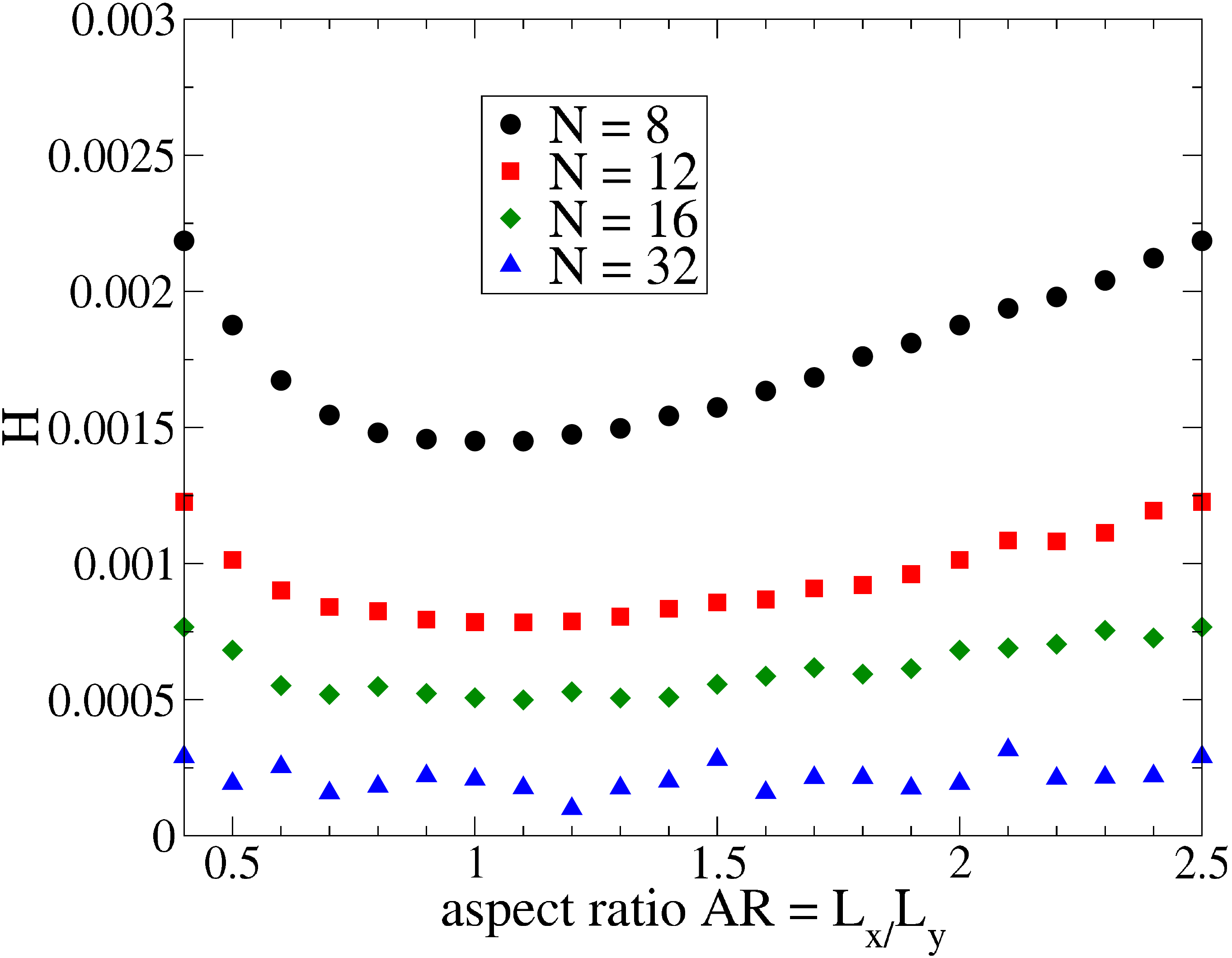}
\end{center}
\caption{\label{fig:acbb}
  (a) Values of $A-C$ in Eq.~(\ref{HFenergy}) for a torus of $N_\phi=8,12,16,32$.
  The cutoff in momentum space was $|m|,|m'|\le 100N_\phi$.
  The value for an infinite system would be $-\frac{1}{32}\sqrt\frac{\pi}{2}\approx-0.039$.
  (b) The same for $B$ in Eq.~(\ref{HFenergy}).
  This term would vanish for the infinite system.
  (c) The same for $H$ in Eq.~(\ref{HFenergy}).
  This term would vanish for the infinite system.
}
\end{figure}

For a comparison with exact diagonalization, we consider the Hartree-Fock approximation on the torus.
By standard mean-field decoupling of the interaction part of the Hamiltonian
\begin{eqnarray}
  \mathcal H_\textrm{\scriptsize C}&=&\frac{1}{2}\sum_{n_1,n_2,n_3,n_4=0}\sum_{m_1,m_2,m_3,m_4=0}^{N_\phi-1}
  \mathcal{A}_{m_1,m_2,m_3,m_4}^{n_1,n_2,n_3,n_4}
  :\rho_{n_1m_1,n_4m_4}\rho_{n_2m_2,n_3m_3}:,\\
  \rho_{nm,n'm'}&=&c^\dag_{nm} c_{n'm'},
\end{eqnarray}
where the matrix elements $\mathcal{A}_{m_1,m_2,m_3,m_4}^{n_1,n_2,n_3,n_4}$ among the basis states $\eta_{nm}$ are given in
Eq.~(\ref{eq:matrixelement}) below.
The uniform state on the torus is the one where each guiding center position is occupied by
the same linear combination of the $\LLletter=0,1$ orbitals~:
\begin{equation}
  \label{eq:filled}
  \left\vert \Psi_\textrm{\scriptsize U}\right\rangle=
  \prod_{m=0}^{N_\phi-1}\left(\cos\left(\frac{\theta}{2}\right)e^{i\phi/2}
  c^\dag_{0m} + \sin\left(\frac{\theta}{2}\right)e^{-i\phi/2}c^\dag_{1m} \right)|0\rangle.
\end{equation}
With the definition~:
$\rho^x_{mm'}=(\left\langle\rho_{1m,0m'}\right\rangle + \left\langle\rho_{0m,1m'}\right\rangle)/2$,
$\rho^y_{mm'}=(\left\langle\rho_{1m,0m'}\right\rangle - \left\langle\rho_{0m,1m'}\right\rangle)/2i$,
$\rho^z_{mm'}=(\left\langle\rho_{0m,0m'}\right\rangle - \left\langle\rho_{1m,1m'}\right\rangle)/2$,
this state corresponds to the dipole density~:
\begin{equation}
  \vec\rho_{mm'}=\delta_{mm'}\frac{1}{2}\left(\begin{array}{c} \sin\theta\cos\phi \\ \sin\theta\sin\phi \\ \cos\theta\end{array}\right).
\end{equation}
By some algebra, we arrive at the energy expression~:
\begin{equation}
  \frac{E_\textrm{\scriptsize HF}}{N}=
  -\epsilon_{N}
  +\frac{h+H}{2}\cos\theta
  +A\sin^2\theta + C\cos^2\theta + B\sin^2\theta\cos(2\phi),
  \label{HFenergy}
\end{equation}
where $A, B, C, H$ are some constants
and the energy $\epsilon_{N}$ has a lengthy expression that goes to
$-\frac{11}{32}\sqrt{\frac{\pi}{2}}$ for large $N$.
These quantities have the symmetry $A(\aspectratio)-C(\aspectratio)= A(1/\aspectratio) - C(1/\aspectratio)$,
$B(\aspectratio)=-B(1/\aspectratio)$, and $H(\aspectratio)=H(1/\aspectratio)$.
Hence $B$ vanishes for a square torus $\aspectratio=1$, and we also have $B>0\iff\aspectratio<1$.
Thus, for $\aspectratio>1$ we have $\phi={\pi}/{2}$, otherwise $\phi=0$;
if $\aspectratio\to {1}/{\aspectratio}$, the preferential direction of the pseudospins rotate by ${\pi}/{2}$,
and there is no preferential direction on the square torus.
Numerical values for various $N_\phi$ are shown in Fig.~\ref{fig:acbb}.
Clearly, the deviation from the infinite-system values is a finite size effect that disappears with increasing $N_\phi$.
We note that $H$ is a tiny correction as compared to $h$, even for small systems.

\begin{figure}[htbp]
\begin{center}
\includegraphics[width=0.49\textwidth, keepaspectratio]{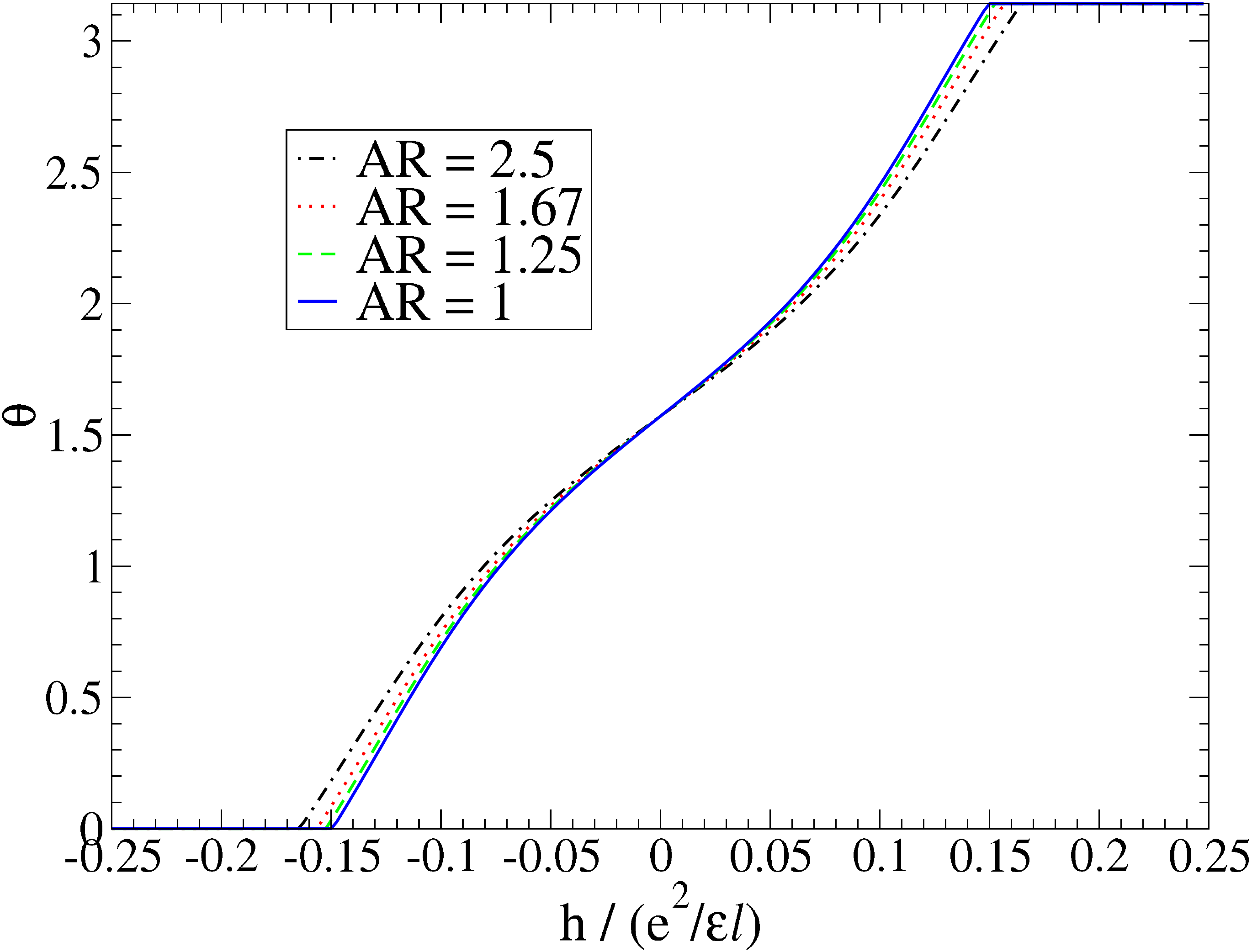}
\includegraphics[width=0.49\textwidth, keepaspectratio]{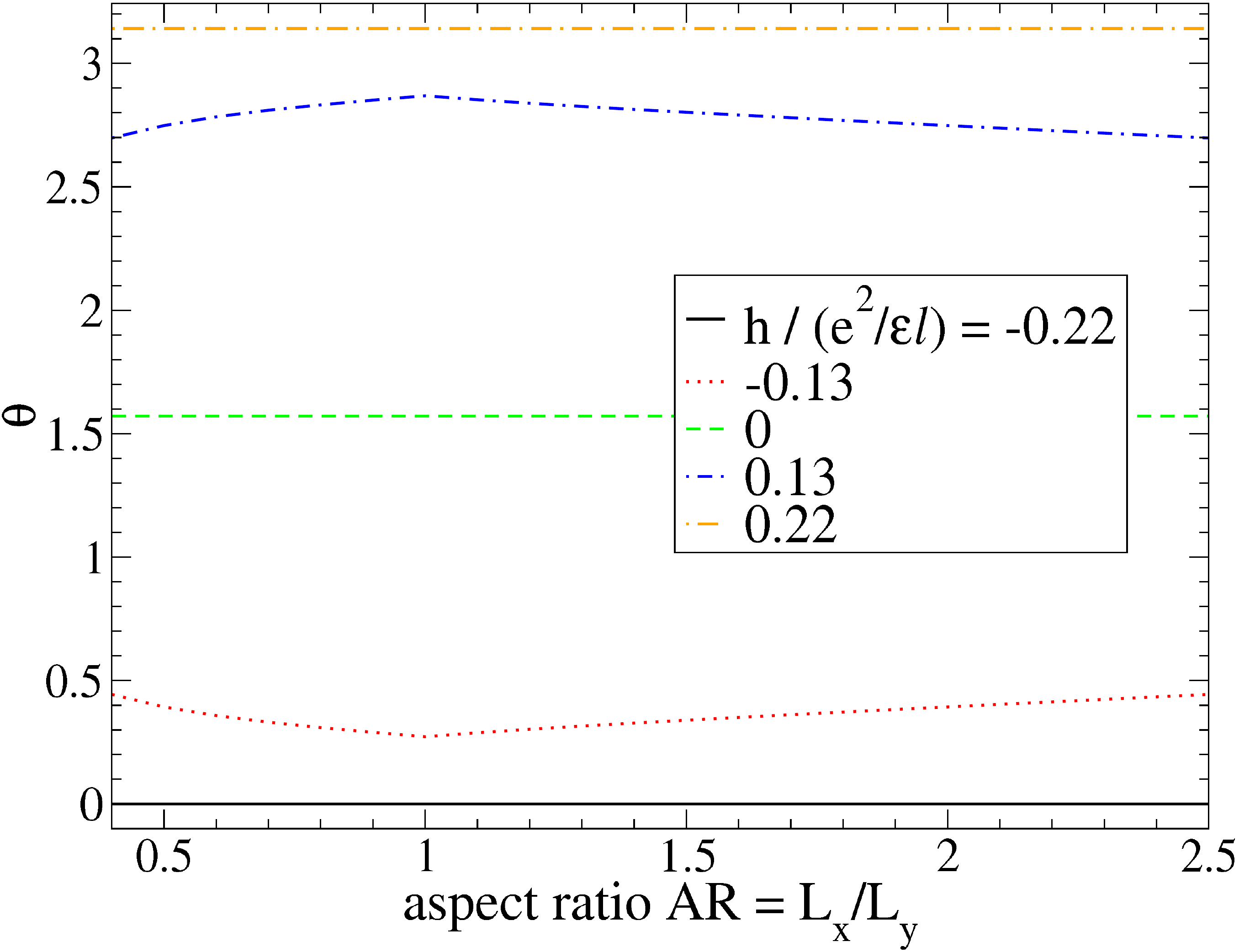}
\end{center}
\caption{\label{fig:theta}
  Left panel: The $\theta$ angle of the uniform state that corresponds to the pseudospin polarization as a function of the bias
  $h$ at fixed aspect ratio for $N_\phi=12$.
  Right panel: $\theta$ as a function of the aspect ratio at fixed bias.
}
\end{figure}

Finally we have~:
\[
  \frac{E_\textrm{\scriptsize HF}}{N\left(\frac{e^2}{\epsilon\ell}\right)}=-\epsilon_N+ 
  \frac{h+H}{2}\cos\theta+(A-|B|)\sin^2\theta + C\cos^2\theta.
\]
This energy is minimized by $\cos\theta=\frac{h+H}{4(A-C-|B|)}$.
Fig.~\ref{fig:theta} shows the $\theta$ angle that corresponds to the pseudospin polarization as a function of the bias
$\frac{h\epsilon\ell}{e^2}$ at fixed aspect ratio, as well as the function of the aspect ratio at fixed bias values.

We conclude that the finite system size hardly changes the Hartree-Fock prediction that the uniform phase
  exhibits orbital coherence in a wide range of the orbital bias, $-0.15\lesssim h\lesssim 0.15$, irrespective
  from the aspect ratio. The deviation of the orbital polarization for system sizes accessible in exact diagonalization
  from the thermodynamic limit is small.


\section{Exact diagonalization results}

We now expose results from our studies using exact diagonalization for a small number of electrons in the torus geometry,
as defined in Sec.~\ref{sec:torus}.
The matrix elements of the Coulomb interaction among the single-body states in Eqs.~(\ref{eq:eta0}) and (\ref{eq:eta1}) are
\be
\label{eq:matrixelement}
\mathcal{A}_{j_1,j_2,j_3,j_4}^{n_1,n_2,n_3,n_4}=
\frac{1}{L_x L_y}\sum_{n,m}^\prime\,\,
\frac{2\pi e^2}{\epsilon q}\,\,
{\mathrm{e}}^{-\frac{1}{2}(q_x^2+q_y^2)}\,\,
{\mathrm{e}}^{2i\pi \frac{n}{N_\phi}(j_1-j_3)}\,\,
F_{n_1,n_4}(q_x,q_y)F_{n_2,n_3}(-q_x,-q_y)
\delta^\prime_{j_1+j_2,j_3+j_4}\,\,
\delta^\prime_{m,j_1-j_4}
\ee
where the sum over $n,m$ also includes momenta given by $q_x=2\pi n/L_x,q_y=2\pi m/L_y$,
and $\delta^\prime_{n,m}$ stands for Kronecker delta modulo $N_\phi$.
The prime over the double sum means that we omit the $n=m=0$ term i.e.
the $q_x=q_y=0$ contribution of the Coulomb potential due to the
neutralizing background. The form factors $F_{n_1n_2}(q_x,q_y)$ encapsulate the dependence upon orbital index~$n_i$~:
\be 
F_{00}=1, \quad F_{11}(q_x,q_y)=1-\frac{1}{2}(q_x^2+q_y^2),
\ee
\be
F_{01}=-\frac{\ell}{\sqrt{2}}(iq_x+q_y),\quad F_{10}=-\frac{\ell}{\sqrt{2}}(iq_x-q_y).
\ee
Static effects of screening can be taken into account by multiplying the Coulomb potential
$2\pi e^2/q$ by an appropriate function.
The Hamiltonian we diagonalize is thus given by~:
\be
\mathcal{H}=
\frac{1}{2}\sum_{n_i,j_i}\mathcal{A}_{j_1,j_2,j_3,j_4}^{n_1,n_2,n_3,n_4}
c^\dag_{n_1,j_1}c^\dag_{n_2,j_2}c_{n_3,j_3}c_{n_4,j_4}
+hN_1
\label{Htot}
\ee
where the pseudomagnetic field $h$ parameterize the bare splitting between $\LLletter=0$ and $\LLletter=1$ Landau levels.

If we consider the particle-hole symmetry then the Hamiltonian Eq.(\ref{Htot})
undergoes the change of field~:
\be
+hN_1 \rightarrow - (h+\Delta_0+\Delta_1) N_1,
\ee
which implies that particle-hole symmetry requires the change $h\rightarrow h_c-h$ with a special critical field $h_c=-1/2(\Delta_0+\Delta_1)$ which is \textit{non zero}.

The algebra of magnetic translations can be used by build many-body conserved momenta~\cite{FDM},
allowing to factor out the degeneracy due to the center of mass momentum.
If we have $N_e$ electrons at flux $N_\phi$ then one defines $N\equiv N_e/N_\phi$ the GCD when the filling 
fraction is $p/q$ i.e. $N_e /N_\phi \equiv pN/qN$ and $p$ and $q$ are coprime.  There are two conserved momenta $K_x, K_y$ that
are living in a Brillouin zone and are discrete. We define two conserved 
\textbf{integer} quantum numbers $s,t$ by~:
\be
\mathrm{K}_x=\frac{2\pi}{L_x}s\, , \quad \mathrm{K}_y=\frac{2\pi}{L_y}t
\ee
where $s,t$ can be taken to vary in the interval $(0,N-1)$ because there is a 
period $N$ i.e. $s+N\equiv s$, $t+N\equiv t$. The modulus of the many-body 
momentum vector is then given by~:
\be
K=\sqrt{K_x^2+K_y^2}=\sqrt{\frac{2\pi}{N_\phi}}\sqrt{\frac{s^2}{\aspectratio} 
+t^2\times \aspectratio}.
\ee
One can now plot the eigenstates vs the total momentum $K$.
The quantum numbers $s,t$ form a grid of $N^2$ points but in fact due to 
discrete symmetries many of them are symmetry-related. Notably there are 
symmetries with respect to $s\rightarrow N-s$ and $t\rightarrow N-t$ that are 
the discrete symmetries of the rectangular unit cell. As a consequence it is 
enough to get spectra in the region $s,t=0,\dots N/2$ since reflections lead 
then to all states. 

There is a subtlety to note~: the origin of the quantum numbers $s,t$ is not 
always zero. In fact when $pq(N_e-1)=2k+1$ then the origin should be taken as 
$s_0,t_0=N/2$ while otherwise, $pq(N_e-1)=2k$ the origin is zero. It means that 
one has to include a shift of N/2 in the momentum definition Eq.(2).
This is important in practice but has no other consequences.

The way we analyze numerical results is as follows~: the nature of the phase of the system
is set by the value of the control parameters in the projected Hamiltonian. In our case
it is the pseudofield $h$ and any kind of screening parameter modifying the Coulomb interaction. For fixed parameters we put the system on a rectangle whose aspect ratio
can be chosen at will. This change will not tune the nature of the phase. However some phases may be revealed more easily for some special aspect ratio. In the case of incompressible fluid states exhibiting the FQHE it is known that they are very insensitive
to the geometry as expected for a liquid state of matter. On the contrary states which break translation symmetries are very sensitive to the aspect ratio. It is by fine tuning that one can observe a set of quasi-degenerate states that are the hallmark of broken translation symmetry~\cite{RHY,HRY,YHR}. This happens in the case of the Wigner crystal state at low filling factors in the LLL and also in the case of stripe states observed at half-filling in high enough LLs. When the aspect ratio is not tuned to the optimal value then the spectrum is featureless as expected for a compressible system.

\section{Polarized incoherent phase}

\begin{figure}[t]
\centering
 \includegraphics[width=0.5\columnwidth]{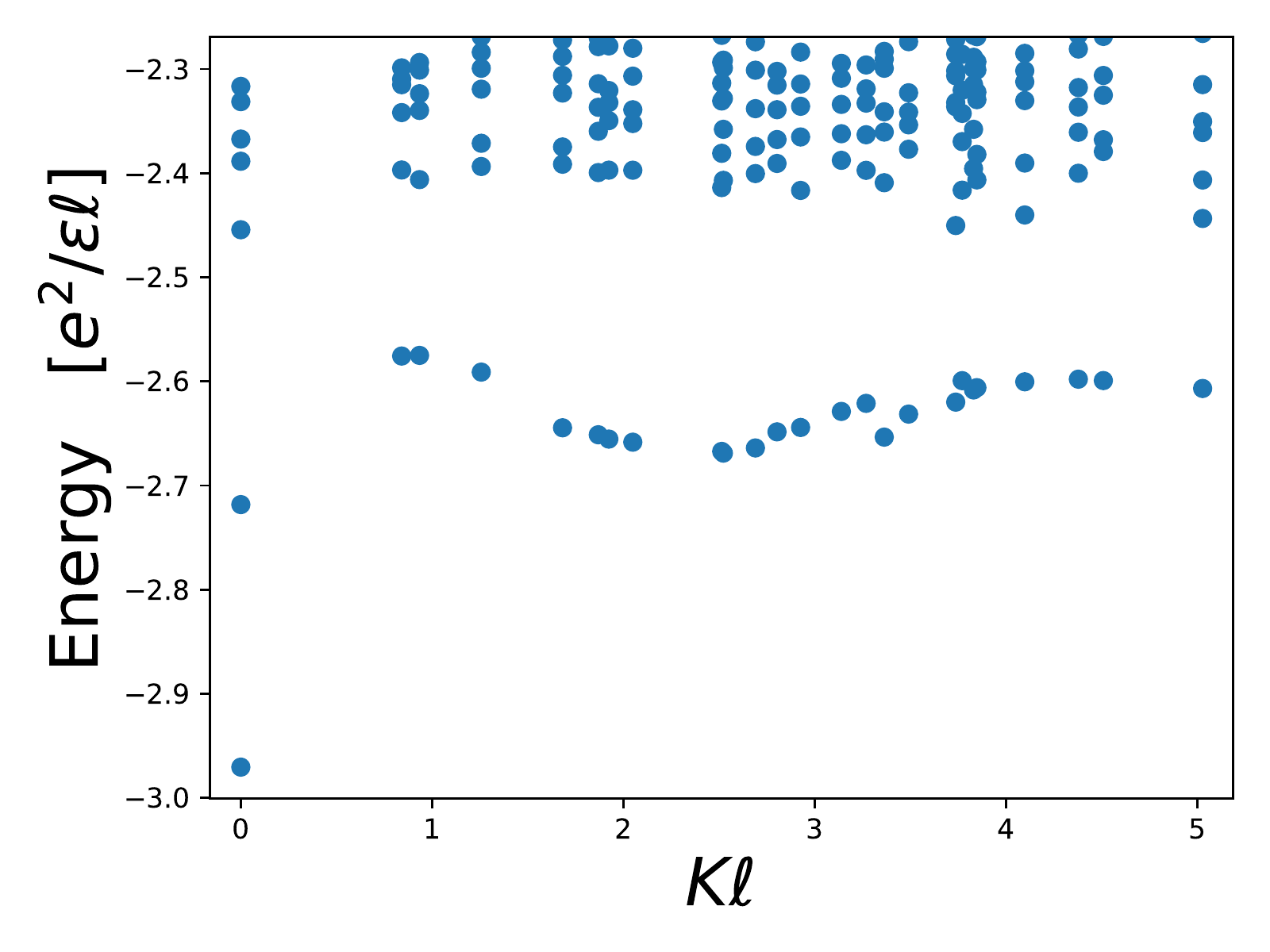}
 \caption{Exact eigenstates of a $N_e=8$ electron system at $\nu=1$ when the bare
 splitting h favors the accumulation into the LLL. The first excited states are expected to be
 magneto-excitons obtained by promoting a single electron into the $\LLletter=1$ LL. This consistent with the well-defined excitation branch above the $K=0$ ground state. Here the aspect ratio is tuned to $\aspectratio=1$. The branch is flatter when we increase the $\LLletter=0$/$\LLletter=1$ splitting given by the field $h$.}
 \label{triv-phase}
\end{figure}

\begin{figure}[t]
\centering
 \includegraphics[width=0.5\columnwidth]{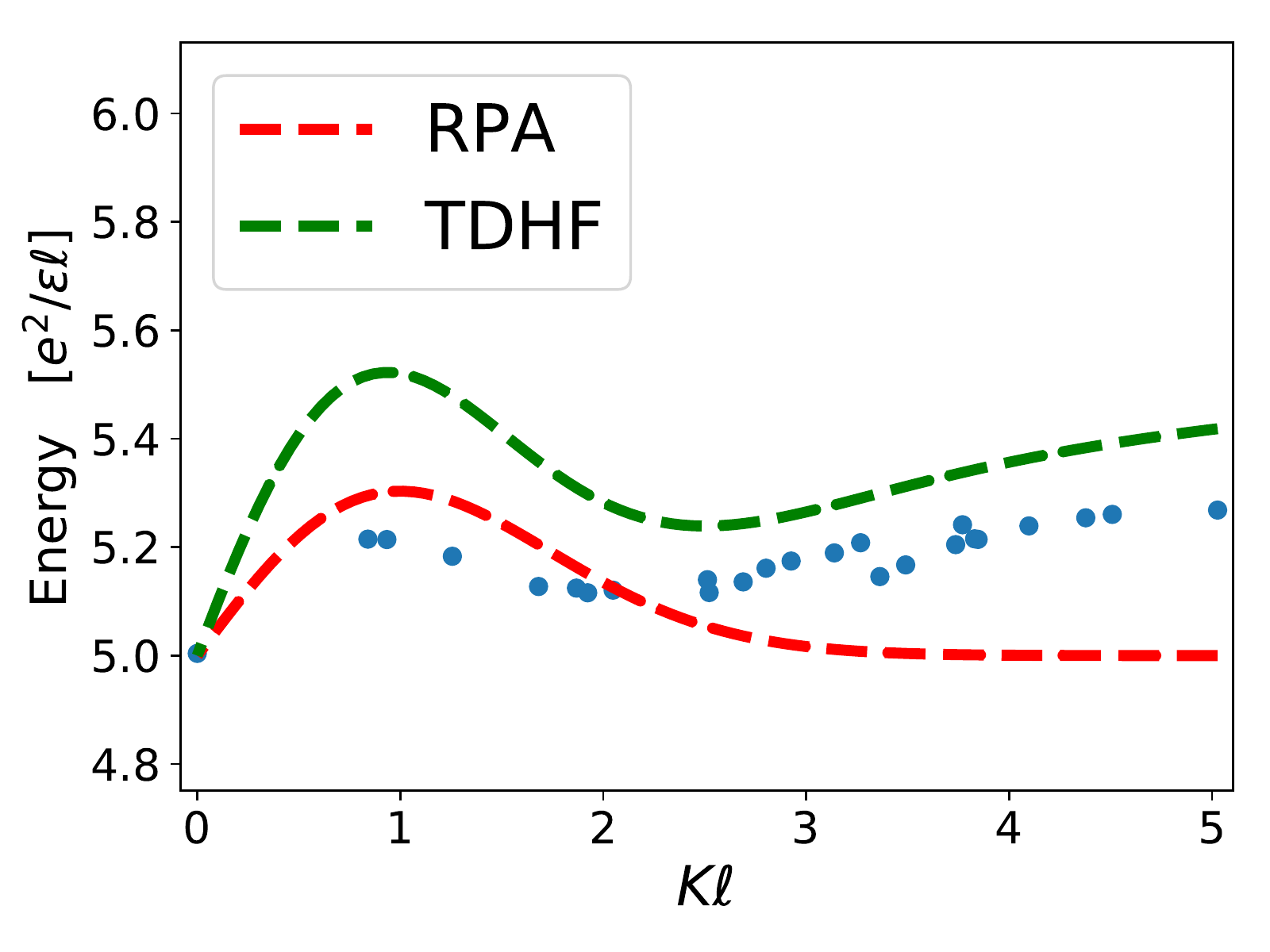}
 \caption{Dispersion of the collective mode in the incoherent phase. Here we have $N_e=8$ electrons at aspect ratio 0.9 and the level splitting is taken to be $h/(e^2/(\epsilon\ell))=5$. The ED points are blue discrete points while RPA approximation to the magneto-exciton is the red line and the TDHF result is green.}
 \label{disp5}
\end{figure}

When the splitting between the two $\LLletter=0/\LLletter=1$ Landau levels is very large, then the physics becomes simple. If for example the $\LLletter=1$ states are at very large energy
then the $\nu=1$ ground state is obtained by filling all orbitals of the $\LLletter=0$ levels.
For  $\nu=1$ there is only one way to do it and this leads to a single Slater determinant.
As a consequence this state is uniform and has total momentum $K=0$.
Excited states are also simple. The first excited state is obtained by promoting only one electron
in the upper $\LLletter=1$ band. As can be observed in Figure (\ref{triv-phase}) there is a well-defined collective mode that we interpret as the magnetoexciton.
This magnetoexciton has a dispersion that can be computed 
by standard analytical  techniques. Within random-phase approximation the dispersion 
relation is given by the zeroes of $1-v_q \chi_0(q,\omega)$ where $v_q=2\pi e^2/\epsilon q$ is the Fourier transform of the Coulomb potential and $\chi_0(q,\omega)$ the so-called Lindhard function~:
\be
\chi_0(q,\omega) \equiv
\frac{1}{2\pi}\sum_{j,j^\prime}|F_{jj\prime}(q)|^2
\frac{n(\varepsilon_j)-n(\varepsilon_j^\prime)}{\hbar\omega+(j-j^\prime)h+i\eta}=
\frac{e^{-q^2\ell^2/2}}{2\pi}(\frac{q^2}{2})
\left[
\frac{1}{\hbar\omega-h+i\eta}-
\frac{1}{\hbar\omega+h+i\eta}
\right] ,
\ee
where $n(\varepsilon)$ is the occupation number and $\eta$ a vanishing regulator.
This leads to the RPA formula~:
\be
\omega^{RPA}_q=h+(\frac{e^2}{\epsilon\ell})\frac{1}{2}q\ell \, e^{-q^2\ell^2/2}
\label{RPA}
\ee
It is also possible to derive the time-dependent Hartree-Fock dispersion relation
following the treatment of full Landau levels~\cite{GiulianiVignale}~:
\be
\omega^{TDHF}_q=h+(\frac{e^2}{\epsilon\ell})q\ell\, e^{-q^2\ell^2/2}
+(\frac{e^2}{\epsilon\ell})\int_0^{\infty}dx\, e^{-x^2/2}(1-L_1(\frac{x^2}{2}))
-(\frac{e^2}{\epsilon\ell})\int \frac{d\mathbf{w}}{2\pi}
\frac{e^{-w^2/2}}{|\mathbf{w}+\mathbf{q}\ell\times\mathbf{{\hat z}}|}
\frac{w^2}{2}
\label{TDHF}
\ee
which leads to the following explicit formula~:
\be
\omega^{TDHF}_q=h+(\frac{e^2}{\epsilon\ell})q \ell\, e^{-q^2\ell^2/2}
-(\frac{e^2}{\epsilon\ell})\frac{1}{4}\sqrt{\frac{\pi}{2}}
e^{-q^2\ell^2/4}q^2\ell^2 \left[ I_0(\frac{q^2\ell^2}{4})-I_1(\frac{q^2\ell^2}{4})\right]
+(\frac{e^2}{\epsilon\ell})\frac{1}{2}\sqrt{\frac{\pi}{2}}
\left[1- e^{-q^2\ell^2/4}I_0(\frac{q^2\ell^2}{4})\right]
\ee
These two dispersion relations tend to $\omega=h$ when $q\rightarrow0$
which a remnant of Kohn theorem in this system restricted to two LLs. However we note that the behavior at large momentum is very different between RPA and TDHF. While $\omega^{RPA}_q \rightarrow h$
for $q\rightarrow\infty$ the TDHF result goes to a finite limit greater than $h$.
Indeed for large momentum the separation between the electron and its associated hole
depends on the quantity $\mathbf{q}\ell\times\mathbf{{\hat z}}$ which is the separation
between them in units of $\ell$ so for large $q$ the electron and hole are far apart and 
the energy of the magneto-exciton approaches the sum of the HF energies of the two particles plus their Coulomb interaction $-e^2/(\epsilon q\ell^2)$. This is behavior is observed in our ED studies. In Fig.(\ref{disp5})  for $h=5\times e^2/(\epsilon\ell)$
we see that the RPA fails to reproduce the large momentum behavior while the TDHF approximation is in good agreement with ED results. However this is true only in the large
splitting $h$ limit. 

The HF theory thus predicts that the incoherent polarized state would become unstable for $h<0$. We find instead that the incoherent phase is stable up to negative values of the bare field
$h_{c1}\approx -0.1 e^2/(\epsilon\ell)$ and also by use of the particle-hole symmetry discussed in section III it extends also from the symmetric field from $h_c=-\infty$ to $h_{c2}\approx -0.2$. This means that the special case with $h=0$ is in fact within
the incoherent phase, in agreement with the findings of Ref.~(\onlinecite{AbaninPapic}). The excitation spectrum is just a distortion of that
in Fig.(\ref{triv-phase}).
A hypothetical uniform orbitally coherent phase would be accompanied by a Goldstone mode~\cite{Barlas08,Barlas10,Cote10,Cote11}. We do not observe such a softening of the low-lying excitations. Instead, as we will see in the next section, we find evidence for a phase
that breaks spontaneously translation symmetry.

\section{Helical phase}

\begin{figure}[t]
\centering
 \includegraphics[width=0.5\columnwidth]{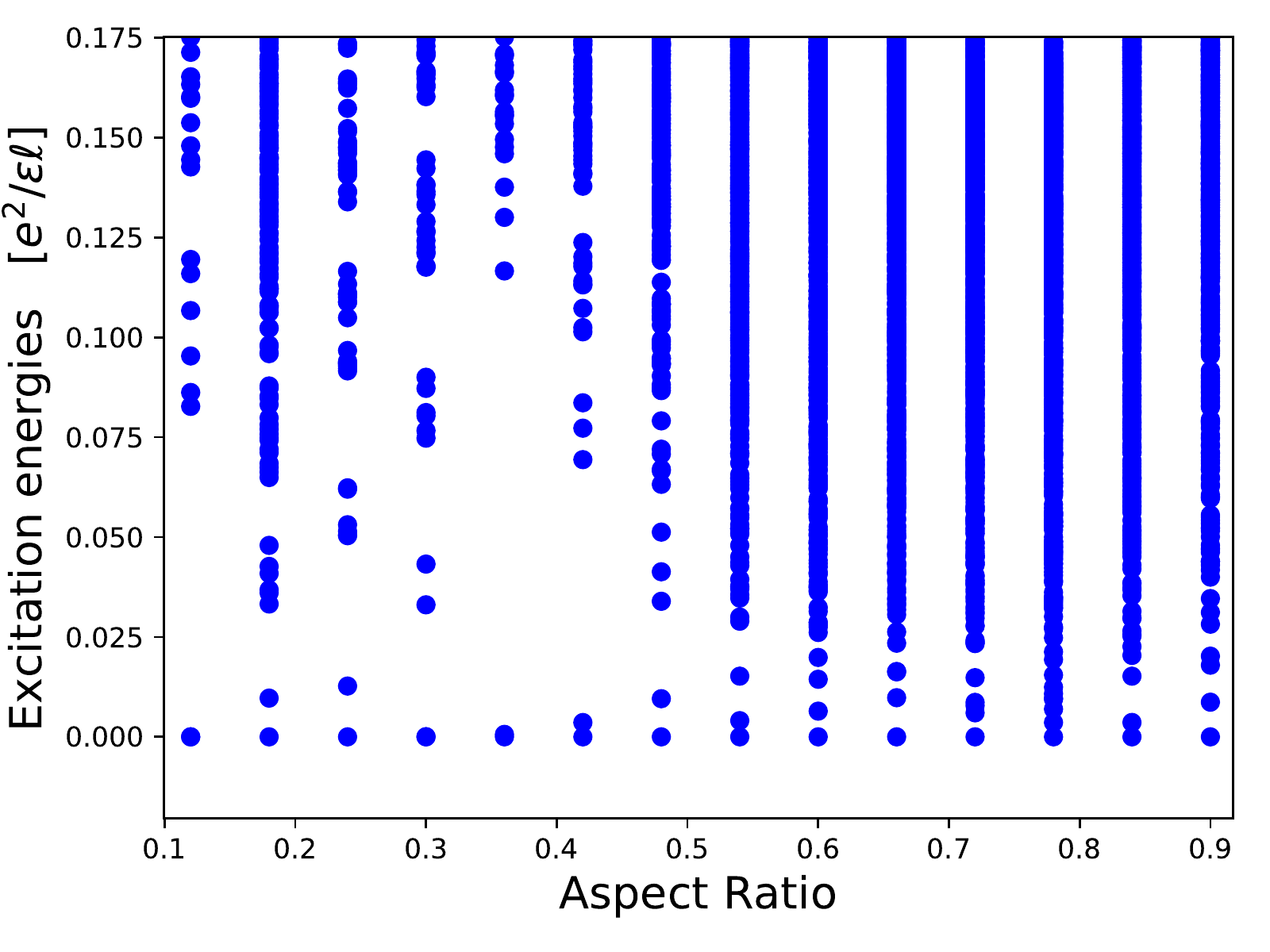}
 \caption{Excitation gap above the $K=0$ ground state of $N_e=10$ electrons as a function of the aspect ratio. 
 The pseudomagnetic field is tuned at the particle-hole invariant value which is in the middle of the \textbf{helical} phase. 
 The ground state degeneracy is clear when the aspect ratio is close to 0.35.}
 \label{ARscan}
\end{figure}

\begin{figure}[t]
\centering
 \includegraphics[width=0.5\columnwidth]{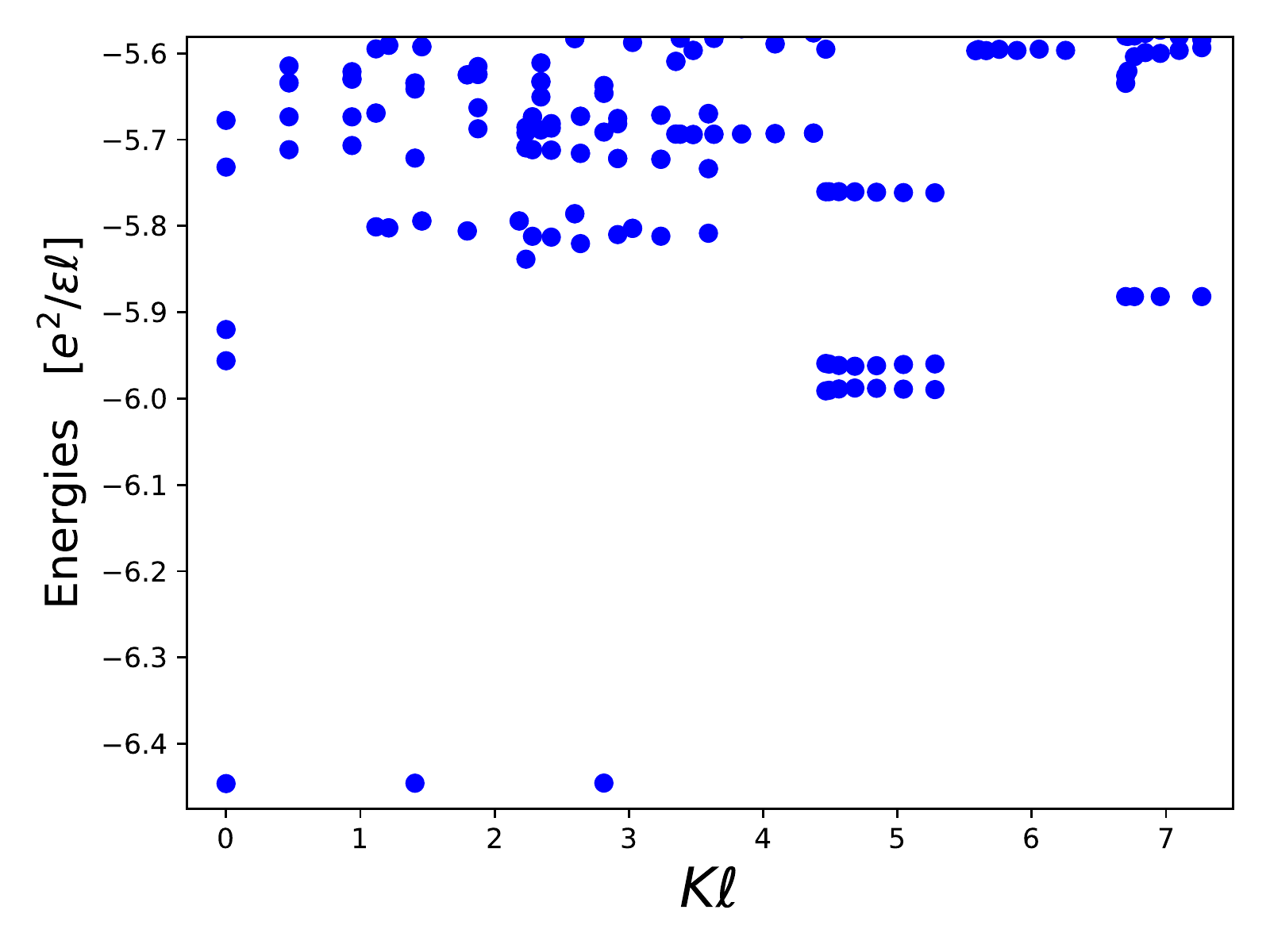}
 \caption{Spectrum of $N_e=12$ electrons for aspect ratio 0.45. The pseudomagnetic field is tuned at the particle-hole invariant value which is in the middle of the \textbf{helical} phase. The ground state degeneracy is evidence for a state with spontaneously broken translation symmetry.}
 \label{stripe}
\end{figure}

The incoherent phase is quite insensitive to the aspect ratio of the rectangular system.
If now we tune the field $h$ to negative values we observe a change of behavior. The ground state
is no longer at zero momentum and its location in the Brillouin zone depends on the aspect ratio. 
We have varied the aspect ratio to find the characteristic behavior of the system.
Typical spectra are displayed in Fig.(\ref{ARscan}). There is a range of aspect ratio
close to 0.3-0.4 for which the ground state becomes degenerate and excited states appear clearly above 
as a discrete set of states. These states have many-body momenta that differ by a one-dimensional 
wavevector~: see Fig.(\ref{stripe}). This is the behavior expected from a quantum system that breaks 
a symmetry as observed previously in quantum Hall systems~\cite{HRY,RHY,YHR}. There is a characteristic 
wavevector associated from this manifold of states which is the momentum difference of the low-lying states. 
It is important to note that the region in aspect ratio where this degeneracy appears is not 
adiabatically connected to the thin-torus of AR going to zero.
There are many level crossings before this limit so the helical phase we observe is not a simple 
electrostatic limit of the quantum Hall problem but a real nontrivial many-body state.
We have also checked that these spectral features are stable if we distort the torus into an oblique cell.

\begin{figure}[t]
\centering
 \includegraphics[width=0.5\columnwidth]{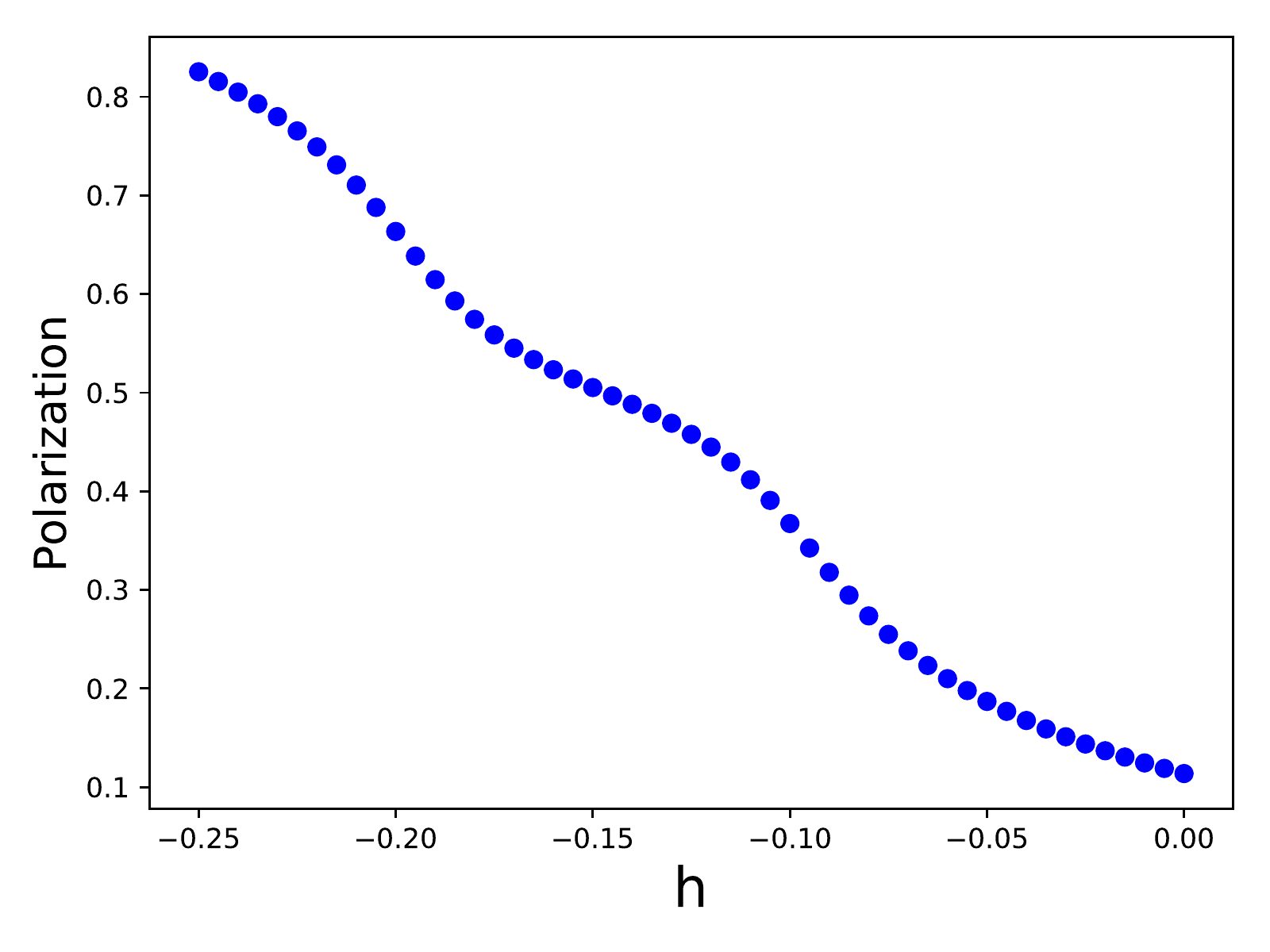}
 \caption{The polarization of the ground state as a function of the field $h$ for aspect ratio 0.9 and 10 electrons.}
 \label{polar}
\end{figure}

To pin down the phase boundary of the helical phase we have used two indicators. The first one is the value of the orbital polarization of the ground state in the $K=0$ sector. 
We define it as the normalized population of the $\LLletter=1$ orbitals:
\be
\mathcal{P}=\langle\Psi|N_1|\Psi\rangle / N_e,
\label{polarization}
\ee
so that $0\leq\mathcal{P}\leq1$ and the polarization goes to zero for large positive $h$ and goes to unity
for large negative $h$. We expect that by the particle-hole symmetry the curve
has a center of symmetry for a non-trivial value of the field. This is exactly what we observe. See Fig.(\ref{polar}) for the polarization of 10 electrons in a rectangle of aspect ratio 0.9. Here we have used the polarization of the ground state in the $K=0$ sector.
This does not mean that this state remains the strict absolute ground state for all values of $h$ and also
for various values of the aspect ratio. Note also that in the incoherent phases for large $h$ values the polarization is either small or close to unity but remains nontrivial as expected from conventional perturbative considerations.

\begin{figure}[t]
\centering
 \includegraphics[width=0.5\columnwidth]{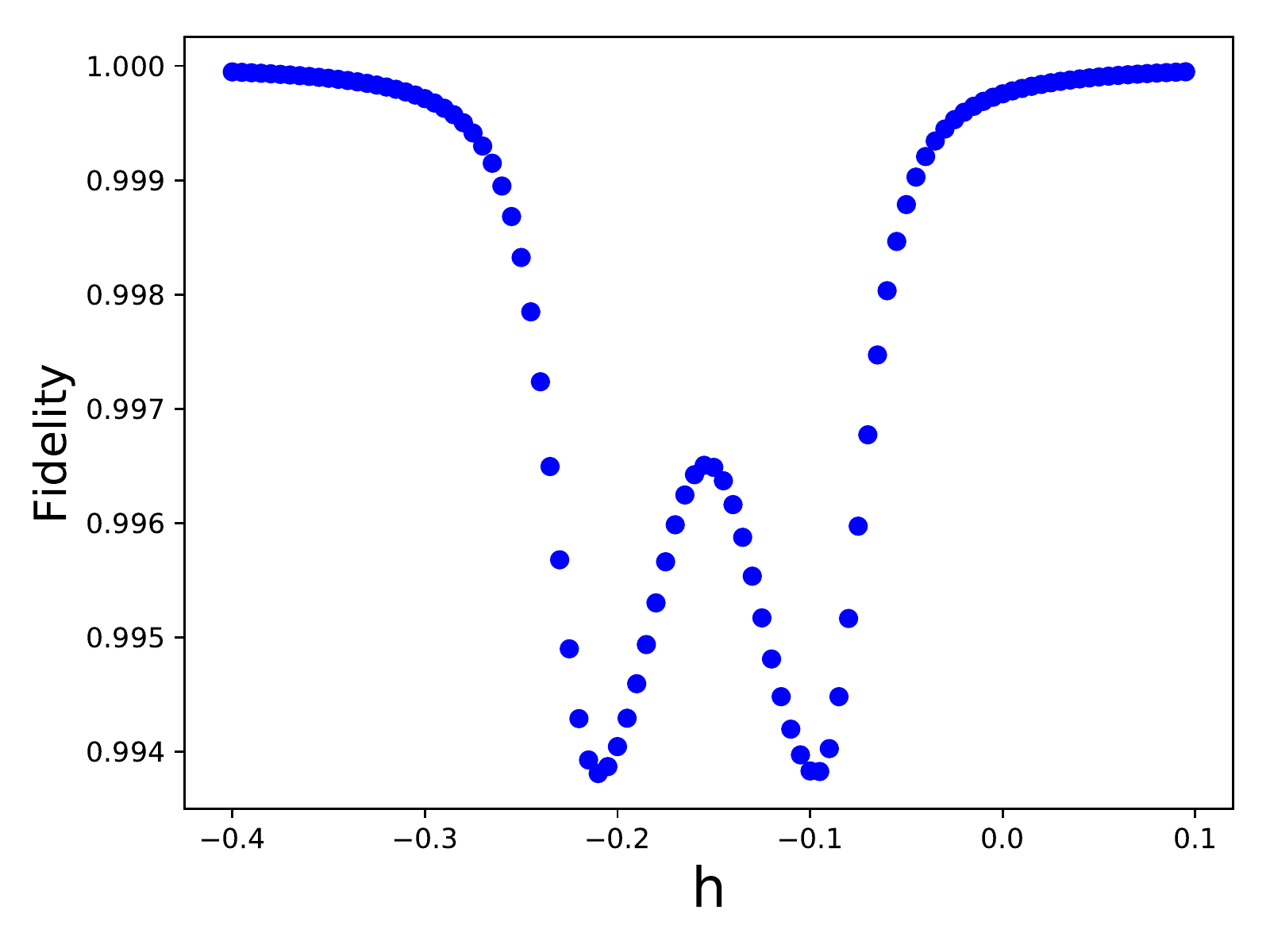}
 \caption{The quantum fidelity $\mathcal{F}$ of the $K=0$ ground state as a function of the field $h$ for aspect ratio 0.9 and 14 electrons.}
 \label{fidelity}
\end{figure}

The other indicator for phase transitions
is the quantum fidelity of the ground state. This is computed by changing the $h$ field
by a small value and computing the overlap between the ground states for these
neighboring values~:
\be
\mathcal{F}=\langle\Psi(h)|\Psi(h+\epsilon)\rangle .
\label{fidelityformula}
\ee
We have computed the fidelity as a function of $h$ for an aspect ratio fixed to 0.9.
It varies very weakly in the fully polarized phase and has strong variations only at what
we define as the helical state. See Fig.(\ref{fidelity}) where we display data for $N=14$ electrons.
The apparent two-peak structure allows us to define finite-system critical values by taking the $h$ values where the fidelity has peaks. This is an alternate way
to characterize phase boundaries of the helical state.
We do not have access to enough system sizes to perform a meaningful finite-size scaling of the fidelity.

\begin{figure}[t]
\centering
 \includegraphics[width=0.5\columnwidth]{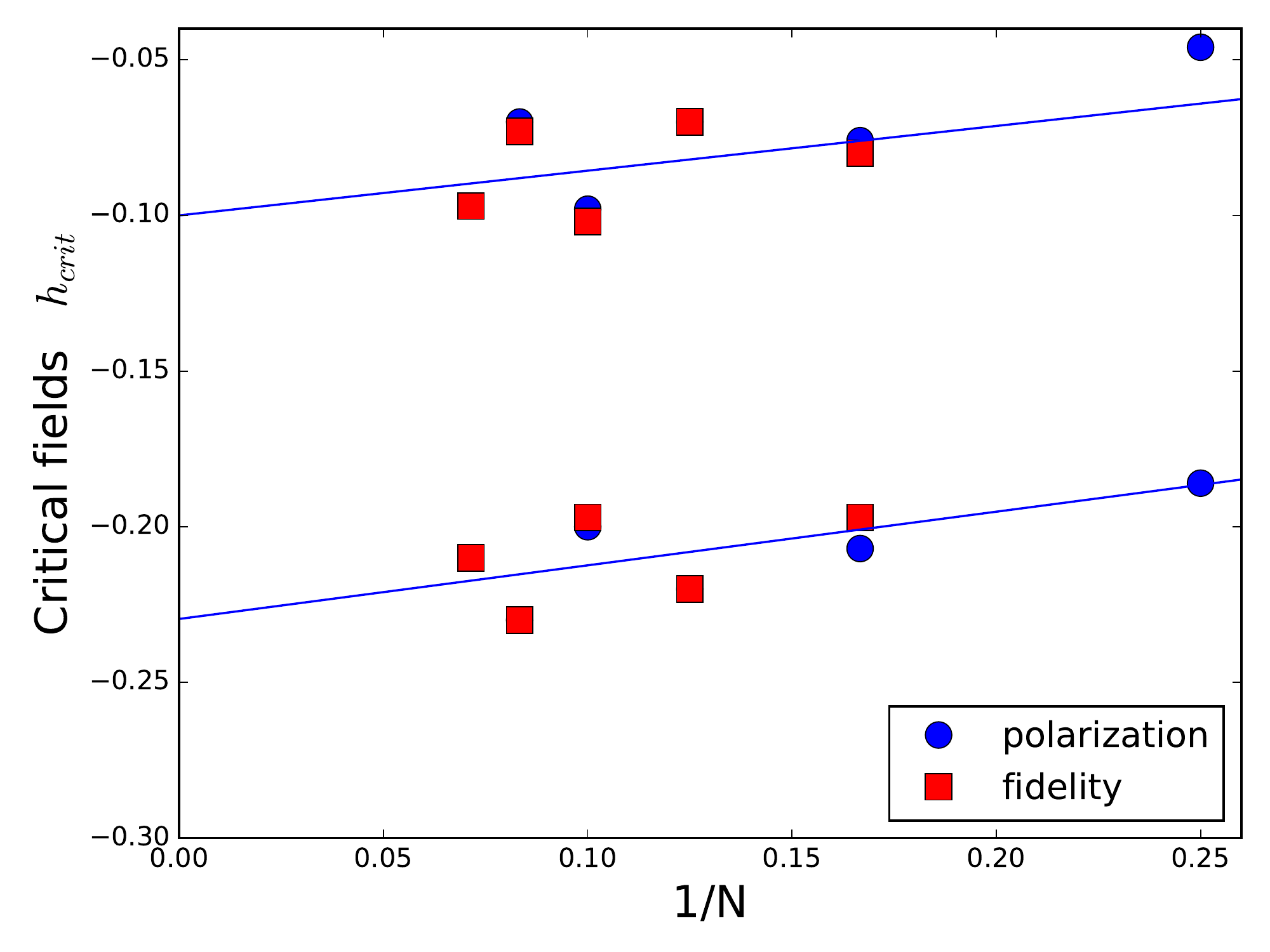}
 \caption{The two critical field delimiting the stability region of the helical phase.
 The values are plotted as a function of the inverse system size.
 The blue circles are obtained from the maximum derivative of the polarization defined in Eq.(\ref{polarization})
 and the red squares from the peak values of the fidelity defined in Eq.(\ref{fidelityformula})}
 \label{bounds}
\end{figure}

The boundaries of the helical phase can thus be pinned down either by looking at the  maximum value of the derivative of the 
polarization Eq.(\ref{polarization}) or also at peak fidelity Eq.(\ref{fidelityformula}). 
These two estimators are plotted in Fig.(\ref{bounds})
as a function of the inverse system size. By use of a linear fit we obtain $h_{c1}\approx -0.1$ and $h_{c2}\approx-0.22$.  
Changing the aspect ratio slightly changes these values and associated uncertainty is estimated to be of the same order at 
that coming from the extrapolation to the thermodynamic limit.

\section{Conclusions}

By means of exact diagonalization on the torus, we have shown that at filling factor $\nu=1$ a system of 
two Landau levels with orbital character $n=0/1$ exhibits a non-trivial ground state with spontaneously 
broken translational symmetry that is stabilized by a orbital single particle splitting ($h$),  within 
the range $ -0.1\gtrsim  h \gtrsim-0.22$. Here negative $h$ favors polarization into the $n=1$LL. This 
state is consistent with the orbitally coherent ferroelectric helical phase that was identified in previous 
Hartree-Fock studies~\cite{Barlas08,Barlas10,Cote10,Cote11,Lambert,Knothe}. Outside of this range of orbital 
splittings the system has an incoherent ground state that is adiabatically connected to the trivial fully polarized 
states into the $n=0/1$ orbitals depending on the sign of $h$, including the case of $h=0$ previously studied in 
Ref.~(\onlinecite{AbaninPapic}). The low lying excitations of these incoherent states are well captured by TDHF to a precision of a few percent. 

We observe a direct transition from this incoherent state into the broken translational symmetry state. Aside 
from the incoherent and the broken translational symmetry phases phases, we see no evidence for a potential uniform 
state with orbital coherence, and, in particular, we do not observe the appearance of its characteristic Goldstone mode. 
The location of the helical phase is in rough agreement with HF predictions, whose boundaries we pin using two criteria: 
quantum fidelity and polarization change. Therefore, we conclude that while HF correctly predicts a helical phase it 
overestimates the tendency to uniform orbital coherent state.

Using a model that includes all the single particle splittings of BLG, we estimate that the orbital splittings can be 
effectively tuned by applying an interlayer electric bias. Our estimates indicate that there is a realistic prospect to 
achieve the regime in which the ferroelectric helical phase becomes the ground state in current experiments~\cite{Hunt,Andrea2017,Dean} 
by applying a large interlayer bias on the order of $u\sim 80{\rm meV}$.

\begin{acknowledgments}
  We acknowledge discussions with Allan MacDonald, Rohit Hegde, Fengcheng Wu.
  We thank IDRIS-CNRS Project 100383 for providing computer time allocation.
  We thank also TOPMAT workshop of PSI2 projet funded by the IDEX Paris-Saclay ANR-11-IDEX-0003-02 where part of this work has been performed.
  C.~T.\ was supported by the National Research Development and Innovation Office of Hungary
  within the Quantum Technology National Excellence Program (Project No.\ 2017-1.2.1-NKP-2017-00001).
\end{acknowledgments}




\end{document}